\documentclass[vecphys]{svmult}

\usepackage{makeidx}         
\usepackage{graphicx}        
\usepackage{multicol}        
\usepackage[bottom]{footmisc}

\makeindex             

\begin{document}


\title*{Gravitational instability in binary protoplanetary disks}
\author{Lucio Mayer\inst{1}
Alan Boss\inst{2} \and Andrew F. Nelson\inst{3}}
\institute{Institute for Theoretical Physics, University of Zurich, and
Institute for Astronomy, ETH Zurich, Zurich, Switzerland
\texttt{lucio@phys.ethz.ch}
\and Carnegie Institution of Washington, Washington, USA \texttt{boss@dtm.ciw.edu}
\and Los Alamos National Laboratory, USA  \texttt{andy.nelson@lanl.gov}}
\maketitle

\begin{abstract}

We review the models and results of simulations of self-gravitating,
gaseous protoplanetary disks in binary star systems. These models
have been calculated by three different groups with three different
computational methods, two particle-based and one grid-based. We show
that interactions with the companion star can affect the temperature
distribution and structural evolution of disks, and discuss the
implications for giant planet formation by gravitational instability
as well as by core accretion. Complete consensus has not been reached
yet on whether the formation of giant planets is promoted or suppressed
by tidal interactions with a companion star. While systems with binary
separations of order 100 AU or larger appear to behave more or less as in
isolation, systems with smaller separations exhibit an increased or
decreased susceptibility to fragmentation, depending on the details of
thermodynamics, in particular on the inclusion or absence of artificial
viscosity, and on the initial conditions. While code comparisons on
identical problems need to be carried out (some of which are already in
progress) to decide which computer models are more realistic, it is
already clear that relatively close binary systems, with separations of
order 60 AU or less, should provide strong constraints on how giant
planets form in these systems.

\end{abstract}

\section{Introduction}

Gravitational instabilities (GIs) can occur in any region of a gas
disk that becomes sufficiently cool or develops a high enough surface
density. In the nonlinear regime, GIs can produce local and global
spiral waves, self-gravitating turbulence, mass and angular momentum
transport, and disk fragmentation into dense clumps and substructure.
It has been quite some time since the idea was first suggested (Kuiper
1951; Cameron 1978) and revived by Boss (1997, 1998a) that the dense
clumps in a disk fragmented by GIs may become self-gravitating
precursors to gas giant planets. This particular idea for gas giant
planet formation has come to be known as the {\sl disk instability}
theory. The idea is appealing since gravitational instability develops
on very short timescales compared to the accumulation of planetesimals
by gravity and the subsequent accretion of gas by a rocky core, the
conventional two-stage formation giant planet formation theory known
as {\sl core accretion} (see the chapter by Marzari et al.).

The particular emphasis of this review chapter is on how gravitational
instability develops when a protoplanetary disk is not isolated but is
a member of a binary or multiple star system (see the chapter by Prato
\& Weinberger). Indeed such a configuration is likely to be the most
common in the Galaxy: the majority of solar-type stars in the Galaxy
belong to double or multiple stellar systems (Duquennoy \& Mayor 1991;
Eggenberger et al. 2004). Radial velocity surveys have shown that
planets exist in binary or multiple stellar systems where the stars
have separations from $\sim$ 20 to several thousand AU (Eggenberger et
al. 2004; see the chapter by Eggenberger \& Udry). Although the
samples are still small, attempts have been made to compare properties
of planets in single and multiple stellar systems (Patience et al.
2002; Udry et al. 2004). Adaptive optics surveys designed to quantify
the relative frequency of planets in single and multiple systems are
underway (Udry et al. 2004; Chauvin et al. 2006). At least 30\% of
extrasolar planetary systems appear to occur in binary or multiple
star systems (Raghavan et al. 2006). These surveys could offer a new
way to test theories of giant planet formation, provided that
different formation models yield different predictions about the
effects of a stellar companion.

\subsection{Gravitational instabilities}

The parameter that determines whether GIs occur in thin gas disks is 

\begin{equation}
\label{eq:Toomre-Q}
Q = c_s\kappa/\pi G\Sigma,
\end{equation}

\noindent
where $c_s$ is the sound speed, $\kappa$ is the epicyclic frequency at
which a fluid element oscillates when perturbed from circular motion,
$G$ is the gravitational constant, and $\Sigma$ is the surface
density. In a nearly Keplerian disk, $\kappa \approx$ the rotational
angular speed $\Omega$. For axisymmetric (ring-like) disturbances,
disks are stable when $Q > 1$ (Toomre 1964). At high $Q$-values,
pressure, represented by $c_s$ in (1), stabilizes short wavelengths,
and rotation, represented by $\kappa$, stabilizes long wavelengths.
The most unstable wavelength when $Q < 1$ is given by $\lambda_{m}
\approx 2\pi^2G\Sigma/\kappa^2$.

Modern numerical simulations, beginning with Papaloizou \& Savonije
(1991), show that nonaxisymmetric disturbances, which grow as
multi-armed spirals, become unstable for $Q < \sim 1.5$. Because the
instability is both linear and dynamic, small perturbations grow
exponentially on the time scale of a rotation period $P_{rot} =
2\pi/\Omega$. The multi-arm spiral waves that grow have a
predominantly trailing pattern, and several modes can appear
simultaneously (Boss 1998a; Laughlin et al. 1997; Nelson et al. 1998;
Pickett et al. 1998).

Numerical simulations show that, as GIs emerge from the linear regime,
they may either saturate at nonlinear amplitude or grow enough to
fragment the disk. Two major effects control or limit the outcome --
disk thermodynamics and nonlinear mode coupling. At this point, the
disks also develop large surface distortions. As emphasized by Pickett
et al. (1998, 2000, 2003), the vertical structure of the disk plays a
crucial role, both for cooling and for essential aspects of the
dynamics.  As a result, except for isothermal disks, GIs tend to have
large amplitudes at the surface of the disk.

Using second and third-order governing equations for spiral modes and
comparing their results with a full nonlinear hydrodynamics treatment,
Laughlin et al. 1998 studied nonlinear mode coupling in the most
detail. Even if only a single mode initially emerges from the linear
regime, power is quickly distributed over modes with a wide variety of
wavelengths and number of arms, resulting in a self-gravitating
turbulence that permeates the disk. In this {\sl gravitoturbulence},
gravitational torques and even Reynold's stresses may be important
over a wide range of scales (Nelson et al. 1998; Gammie 2001; Lodato
\& Rice 2004; Mej\'ia et al. 2005). 

As the spiral waves grow, they can steepen into shocks that produce
strong localized heating (Pickett et al. 1998, 2000a;  Nelson et al.
2000). Gas is also heated by compression and through net mass
transport due to gravitational torques. The ultimate source of GI
heating is work done by gravity. The subsequent evolution depends on
whether a balance can be reached between heating and the loss of disk
thermal energy by radiative or convective cooling. The notion of a
balance of heating and cooling in the nonlinear regime was described
as early as 1965 by Goldreich \& Lynden-Bell and has been used as a
basis for proposing $\alpha$-treatments for GI-active disks
(Paczy\'nski 1978; Lin \& Pringle 1987). For slow to moderate cooling
rates, numerical experiments verify that thermal self-regulation of
GIs can be achieved (Tomley et al. 1991; Pickett et al. 1998, 2000a,
2003; Nelson et al. 2000; Gammie 2001; Rice et al. 2003b; Lodato \&
Rice 2004, 2005; Mej\'ia et al. 2005; Cai et al. 2006a,b). $Q$ then
hovers near the instability limit, and the nonlinear amplitude is
controlled by the cooling rate. There have been various attempts to
model radiative cooling in self-gravitating disks. For a recent
overview of the different methods appearing in the literature and for
a general discussion of gravitational instability in protoplanetary
disks, we point the reader to Durisen et al. (2007). In this chapter
we will focus on the radiative cooling models that have been used in
the few existing works on binary self-gravitating protoplanetary
disks. 

\subsection{Fragmentation and survival of clumps}

As shown first by Gammie (2001) for local thin-disk calculations and
later confirmed by Rice et al. (2003b) and Mej\'ia et al. (2005) in
full 3D hydro simulations, disks with a fixed cooling time,
$t_{cool}=U/\dot{U}$, fragment for sufficiently fast cooling,
specifically when $t_{cool} \le 3 \Omega^{-1}$, or, equivalently,
$t_{cool} \le $ $P_{rot}/2$. Finite thickness has a slight stabilizing
influence (Rice et al. 2003b; Mayer et al. 2004a). When dealing with
realistic radiative cooling, however, one cannot apply this simple
fragmentation criterion to arbitrary initial disk models. One has to
apply it to the asymptotic phase after nonlinear behavior is
well-developed (Johnson \& Gammie 2003). Cooling times can be much
longer in the asymptotic state than they are initially (Cai et al.
2006a,b; Mej\'ia et al., in preparation). For disks evolved under
isothermal conditions, where a simple cooling time cannot be defined,
local thin-disk calculations show fragmentation when $Q < 1.4$
(Johnson \& Gammie 2003). This is roughly consistent with results from
global simulations (e.g., Boss 2000; Nelson et al. 1998; Pickett et
al. 2000a, 2003; Mayer et al. 2002, 2004a). Figure 1 shows a classic
example of a fragmenting disk.

\begin{figure} 
\centering
\includegraphics[height=10cm]{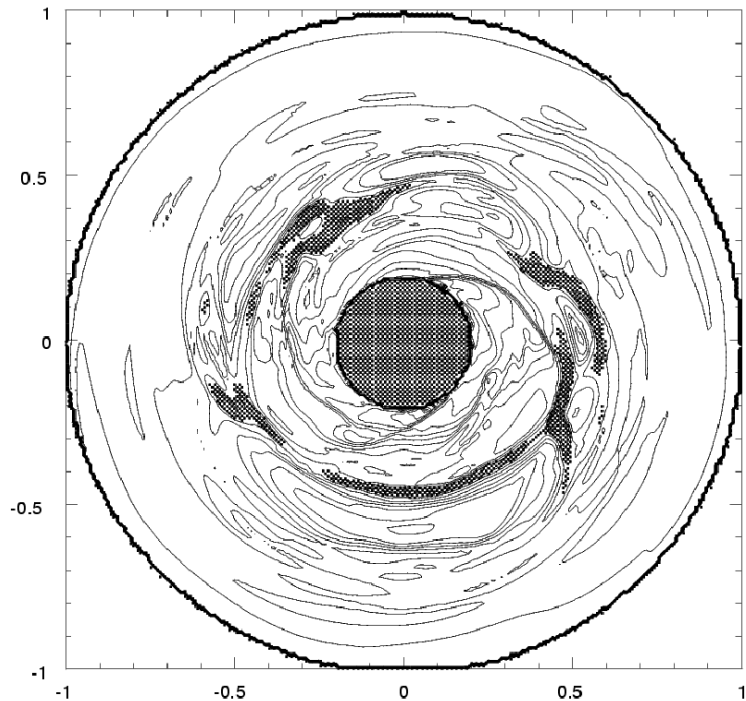}
\caption{Midplane density contours after 339 yr of evolution of a
$0.091 M_\odot$ disk in orbit around a single $1 M_\odot$ protostar, 
showing the formation of a self-gravitating clump of mass
$1.4 M_{Jup}$ at 4 o'clock (Boss 2001).}
\label{fig:1}       
\end{figure}

Rice et al. (2003b) also found that clumps could form in disks with
even longer cooling times ($t_{cool} = 5 \Omega^{-1}$) if the disk
mass was increased. This latter estimate is in agreement with the time
scales for cooling found in 3D models with diffusion approximation
radiative transfer and convective-like motions that led to
fragmentation  into self-gravitating clumps (Boss 2001, 2004a). 

Although there is general agreement on conditions for fragmentation,
two important questions remain. Do real disks ever cool fast enough
for fragmentation to occur, and do the fragments last long enough to
contract into permanent protoplanets before being disrupted by tidal
stresses, shear stresses, physical collisions, and shocks? Recent
simulations have just begun to address the issue of the long term
survival of clumps once they have been produced in a disk (Durisen et
al. 2007). None of these long-term simulations has explored the case
of binary systems and therefore their results are not necessarily
valid in that case (for example they do not take into account the
effect of eventual orbital resonances with the companion that might).
. However, except for clumps forming at the very periphery of one of
the two disks, one would expect the tidal stresses to be dominated by
the central star of their own disk, in which case the results of
isolated disks are still relevant. In addition, clumps are unlikely to
form near the outskirts of disks since the surface density should be
too low there. High spatial resolution appears to be crucial for the
survival of clumps. Pickett et al. (2003) found that 256 azimuthal
cells were not enough to resolve self-gravity on a scale of a fraction
of AU, leading to artificial dissolution of overdensities. An
increased ability of clumps to persist and become gravitationally
bound as the resolution is increased was also found by Boss (2001) and
Mayer et al. (2004). High spatial resolution of the gravitational
force is crucial, as is the accuracy of a gravity solver for a given
resolution element. These features are extremely code-dependent and
are briefly discussed below in section 2.2.

\section{Numerical Techniqes and Assumptions}

To date, only three papers have been published that consider the
possibility of forming giant planets by disk instability in binary
star systems: Nelson (2000), Mayer et al. (2005), and Boss (2006b). In
short, Nelson (2000) found that binarity prevented fragmentation from
occurring, Boss (2006b) found that binarity could enhance clump
formation, and Mayer et al. (2005) found the binarity could discourage
fragmentation in some cases, but permit it in other cases. These three
papers are the focus of the remainder of this chapter, as we try to
decipher the reasons for this apparent dispersion in outcomes.

\subsection{Hydrodynamics methods}

Three codes have been used so far to follow the evolution of binary
protoplanetary disks, two smoothed particle hydrodynamics (SPH) codes
(Nelson 2000; Mayer et al. 2005) and one finite-difference grid-based
code (Boss 2006b; described in detail by Boss \& Myhill 1992). The two
SPH codes are, respectively, a modified version (Nelson 2000) of a
code originally developed by Benz (1990) and GASOLINE (Wadsley et al.
2004; Mayer et al. 2005). We begin with a description of the codes.

Although the two SPH codes are based on a very similar implementation
of SPH, there are differences in some aspects that might be important
for understanding differences in results. One major difference is that
the version of the code used in Nelson (2000) is 2D while GASOLINE is
3D, as is the Boss (2006b) code. Evidence has been accumulated
recently that gravitational instability in a protoplanetary disk is an
intrinsically three-dimensional phenomenon (Cai et al. 2006a,b).
At the same time, at the resolution for which affordable simulations
can be done, 3D codes resolve only very poorly the structure of the
disk in the third dimension. Nelson (2006) has shown that if the
vertical structure is not well resolved, both from a hydrodynamical
standpoint and from a radiative transfer standpoint, serious errors in
the evolution of the disks may develop. Therefore, even with all other
things being equal, this difference alone could result in a different
evolution of the spiral patterns, and thus of the outcome of
gravitational instability. In what follows we will highlight the most
important features of the two SPH codes and we will explicitly
indicate in what ways the two codes differ and what is the expected
outcome of such differences.

SPH is an approach to hydrodynamical modeling first developed by Lucy
(1976) and Gingold \& Monaghan (1977). It is a particle-based method
that does not refer to grids for the calculation of hydrodynamical
quantities: all forces and fluid properties are determined by moving
particles, thereby eliminating numerically diffusive advective terms.
The Boss \& Myhill (1992) code is an Eulerian code, with all
quantities defined on a spherical coordinate grid. The code is
second-order accurate in both space and time, a crucial factor for
keeping numerical diffusion at a tolerable level.

The basis of the SPH method is the Lagrangian representation and
evolution of smoothly varying fluid quantities whose value is only
known at disordered discrete points in space occupied by particles.
Particles are the fundamental resolution elements comparable to cells
in a grid. SPH operates through local summation operations over
particles weighted with a smoothing kernel, $W$, that approximates a
local integral. The smoothing operation provides a basis from which to
obtain derivatives. Thus, estimates of density-related physical
quantities and gradients are generated.

Both GASOLINE and Nelson's code use a fairly standard implementation
of the hydrodynamical equations of motion for SPH (Monaghan 1992). The
density at the location of each particle with index $i$ is calculated
from a sum over particle masses $m_j$

\begin{eqnarray}
\rho_i=\sum_{j=1}^{n} m_j W_{ij}.
\label{denssum}
\end{eqnarray}

\noindent

where $j$ is an index running on the entire set of $n$ particles. The
momentum equation is expressed as

\begin{eqnarray}
\frac{d\vec{v}_i}{dt}& = & -\sum_{j=1}^{n}m_j
\left({\frac{P_i}{\rho_i^2}+\frac{P_j}{\rho_j^2}+\Pi_{ij}}\right)
\nabla_i W_{ij},
\label{sphmom}
\end{eqnarray}

\noindent 
where $P_j$ is pressure, $\vec{v}_i$ is velocity and the 
artificial viscosity term is $\Pi_{ij}$.

The kernel is a standard B-spline with compact support in both codes
(Hernquist \& Katz 1987). The number of neighbors, or in other words,
the number of particles around a given particle that are considered
for smoothed sums, is fixed in GASOLINE at $32$ and varies between 10
and 30 in the 2D version of Nelson's code, depending on the local
flow.

\subsection{Gravity solvers}

Clearly gravity solvers represent a crucial aspect of simulations of
self-gravitating disks. They need to be both accurate and efficient,
and possibly parallelized in order to take advantage of modern
computer architectures and permitting very high resolution
calculations to be performed. Both GASOLINE and Nelson's code compute
gravity using a tree-based solver, which is fast, easily
parallelizable and a natural choice for a particle-based
hydrodynamical such as SPH, since once a tree is built it can also be
re-used as an efficient search method for hydrodynamical forces as
well. Both codes use a modified versions of the binary tree described
in Benz et al. (1990) which approximates the gravity of groups of
distant particles in a multipole expansion while calculating
interactions of nearby particles explicitly. Gravitational forces due
to neighbor particles are softened to avoid divergences as particles
pass near each other. The force calculation in tree algorithms
requires work proportional to $\mathcal{O}(N\,log\,N)$, where $N$ is
the number of particles in the simulation, as opposed to work
proportional to $N^2$ in ``direct'' N-Body algorithms where all
gravitational forces between individual particles are computed
directly. The drawback is that, except for very nearby particles whose
interactions may be calculated as in direct summation codes, the
forces are approximate rather than exact when using a tree. A
particularly useful property of tree codes is the ability to
efficiently calculate forces for a subset of the bodies. This is
critical if there is a large range of time-scales in a simulation and
multiple independent timesteps are employed (see below). At the cost
of force calculations no longer being synchronized among the particles
substantial gains in time-to-solution may be realized. Multiple
timesteps are particularly important for applications where the
primary interest and thus need for high spatial resolution tends to be
focused on small regions within a larger simulated volume; a
protoplanetary disk undergoing fragmentation locally is one such
application. GASOLINE uses 4th (hexadecapole) rather than 2nd
(quadrupole) order multipole moments (as used by most tree codes,
including Nelson's) to represent the mass distribution in cells at
each level of the tree.  This results in less computation for the same
level of accuracy: better pipelining, smaller interaction lists for
each particle and reduced communication demands in parallel. The
current implementation in GASOLINE uses reduced moments that require
only $n+1$ terms to be stored for the $n^{th}$ moment. For a detailed
discussion of the accuracy and efficiency of the tree algorithm as a
function the order of the multipoles used, see Stadel (2001).

Relaxation effects compromise the attempt to model continuous fluids
using particles. Both in GASOLINE and in Nelson's code the particle
masses are effectively smoothed in space using the same spline kernels
employed in the SPH calculation. This means that the gravitational
forces vanish at zero separation and return to Newtonian $1/r^2$ at a
separation of $\epsilon_i+\epsilon_j$ where $\epsilon_i$ is the
gravitational softening applied to each particle. In this sense the
gravitational forces are well matched to the SPH forces. However, in
GASOLINE gravitational softening is constant over time and fixed at
the beginning of the simulation, while in Nelson's code it is
time-dependent and always equal to the local SPH smoothing length.

The use of adaptive softening in Nelson's code ensures that
gravitational and pressure forces are always represented with the same
resolution. Bate \& Burkert (1997) have shown that when an imbalance
between pressure and gravitational forces occurs, artificial
fragmentation or suppression of physical fragmentation can arise.
While Bate \& Burkert (1997) found that such imbalance leads to
spurious results when it occurs at a scale comparable with the local
Jeans length of the system, more recently Nelson (2006) has shown
that particle clumping was amplified for imbalances in
softening/smoothing even when the Toomre wavelength was well resolved,
which is a much more limiting codndition. When the gravitational
softening is fixed over time, such as in GASOLINE, care has to be
taken that this be comparable to the SPH smoothing length at the scale
of the Jeans length. Mayer et al. (2005) choose the softening
according to the latter prescription at the beginning of the
simulation and set its value so that outside $10 AU$ the softening
drops to $\sim 1/2$ the local smoothing length. An argument can be
made that later in the evolution, when strong overdensities develop
along the spiral arms the SPH smoothing length drops significantly,
becoming smaller than the gravitational softening, thereby degrading
the propensity for numerical fragmentation. Both the initial
softening/smoothing inequality and the later evolution of disks
towards fragmentation were examined by Nelson (2006), with the result
that to fragmentation was still enhanced, but after it began,
simulations could be evolved much further, because the fragments did
not continue to contract indefinitely but rather remained comparable
in size to the fixed softening value. No specific tests of whether
this occurs as well in the Mayer et al. work, or how the results may
be affected, has yet been done. It may be of some note however, that
the fragmentation in their work only occured in the outer half of
their disks, where the force imbalances favor gravity, as seen in
Nelson (2006). 

An on going comparison between SPH and adaptive mesh refinement (AMR)
codes conducted at different resolutions and particle softening is in
preparation (Mayer et al., in preparation, see also Durisen et al.
2007) and may provide an independent check of the reliability of
fragmentation in SPH. On the other hand, since this happens when the
gas is optically thick, cooling is also suppressed (see below), and
this might be the dominant, physical effect in slowing down the
collapse. Even so, compensating for insufficiencies in a physical
model, using flaws in a numerical code, would seem to be, at best,
undesirable. Adaptive softening is also not without flaws -- particles
have a time-dependent potential energy, which induces fluctuations in
the potential that can in principle increase force errors, possibly
affecting the accuracy of the integration. This is known to be
problematic for purely gravitational simulations such as those of
cosmic structure formation, but its consequences have not been studied
systematically for the case of self-gravitating fluids. Recent work
of Price \& Monaghan (2006) has shown how adpative softening may be
used while still conserving energy, but their technique has not yet
seen wide adoption.

In the Boss \& Myhill (1992) code, Poisson's equation is solved for
the gravitational potential at each time step. This solution is
achieved by using a spherical harmonic ($Y_{lm}$) expansion of the
density and gravitational potential, with terms in the expansion up to
$l, m =$ 32 or 48 typically being used. Boss (2000, 2001) found that
the number of terms in this expansion was just as important for robust
clump formation as the spatial resolution. Because the computational
effort involved scales as the number of terms squared, however, in
practice this value cannot be increased much beyond 48 without having
the Poisson solver dominate the effort. Boss (2005) showed that the
introduction of point masses to represent very high density clumps led
to better defined, more massive clumps, but the computational effort
associated with introducing these point masses also led to an
appreciable slowing the execution speeds.

\subsection{Timestepping}

Nelson's code uses the Runge-Kutta-Fahlberg method to evolve the
equations of motion. It employs a global timestep for all particles
in the disk, which was increased or decreased depending on the
conditions in the simulations. The integrator is a first order
accurate method with second order error control and the scheme provides
limits on the second order error term in all of the various variables.
Gasoline incorporates the timestep scheme described as Kick-Drift-Kick
(KDK) (see Wadsley et al. 2004). Without gas forces this is a
symplectic leap-frog integration, which ensures the conservation of
total energy, this being not guaranteed with the Runge-Kutta method
(Quinn et al. 1997; Tremaine et al. 2003), though much better
conservation can be had when a single time step for all particles is
used, as was done in Nelson (2000). The leap-frog scheme requires only
one force evaluation and minimum storage. GASOLINE uses multiple
timesteps, hence at any given time different particles in the disk can
be evolved with different timesteps. The base (maximum) timestep is
divided in a hierarchy of smaller steps (rungs), with different
particles being assigned to different rungs. This allows to reach a
much smaller step size when it is required, allowing to probe a much
higher dynamic range and follow correctly the dynamics of regions with
very high densities. The drawback is that the scheme is no longer
strictly symplectic if particles change rungs during the integration
which they generally must do to satisfy their individual timestep
criteria. Adaptivity in the time integration, hence the ability to
achieve small timesteps than the dynamics or hydrodynamics require so,
can be important to model correctly the formation and evolution of
overdensities and clumps. In fact, if the step size is not small
enough the acceleration inside the overdensities, that goes as
$1/{\Delta t}^2$, can be underestimated. This is equivalent to
underestimate the self-gravity of clumps and can in principle lead to
their artificial dissolution, although no systematic tests have ever
been performed. GASOLINE uses a standard timestep criterion based on
the particle acceleration (see Wadsley et al. 2004) and for gas
particles, the Courant condition and the expansion cooling rate.
Nelson's code adopts the Courant condition as well for hydrodynamical
forces and addituonally a set of constraints based on the position and
velocities of particles. If the latter are not met after a timestep,
the scheme went back and tries again with a smaller timestep.

The Boss \& Myhill (1992) code uses a single-size timestep based on
the Courant condition, and uses a predictor-corrector method to
achieve second-order accuracy in time.

\subsection{Artificial viscosity}

Most hydrodynamic methods need artificial viscosity to stabilize the
flow by avoiding particle interpenetration and to resolve in an
approximate manner the physical dissipation present in shocks, and SPH
is no exception. Both Mayer et al. (2005) and Nelson (2000) use bulk
and von~Neumann-Richtmyer (so called `$\bar\alpha$' and `$\beta$')
viscosities to simulate viscous pressures which are linear and
quadratic in the velocity divergence. They both incorporate a switch
(see Balsara 1995) that acts to reduce substantially the large
undesirable shear component associated with the standard form.

In both GASOLINE and Nelson's code the artificial viscosity term reads

\begin{eqnarray}
\Pi_{ij} = \left\{{ \begin{array}{ll}
\frac{-\alpha\frac{1}{2}(c_i+c_j)\mu_{ij}+\beta\mu_{ij}^2}{\frac{1}{2} 
(\rho_{i}+\rho{j})} & {\rm for~}\vec{v}_{ij}\cdot\vec{r}_{ij}< 0,\\
~0 & {\rm otherwise}, \end{array}}\right.\\
{\rm\ where\ }
\mu_{ij} = \frac{h(\vec{v}_{ij}\cdot\vec{r}_{ij})}{\vec{r}_{ij}^{\,2}+0.01 
(h_i+h_j)^2},
\label{artifvisc}
\end{eqnarray}

\noindent 
where $\vec{r}_{ij}=\vec{r}_i-\vec{r}_j$,
$\vec{v}_{ij}=\vec{v}_i-\vec{v}_j$, and $c_j$ is the sound speed. 
$\alpha=1$ and $\beta=2$ are standard values of the coefficients of 
artificial viscosity, known to result in acceptable dissipative
behavior across a wide variety of test problems. 

In protoplanetary disks, Mach numbers are high (of order 10-20)
although shocks are not as strong (i.e. the density jumps are not as
pronounced) as when gas collapses or collides with other gas along
nearly radial trajectories such as in cosmological structure
formation. This is important because it means that the standard
settings for the viscous coefficients may be higher than strictly
necessary for correct evolution of the flow. Since artificial
viscosity is essentially a nuisance with regard to improving the
modeling of the hydrodynamical flow, any improvements which decrease
the importance of unphysical side effects, while retaining the
required stability and dissipative effects of the code, are desirable. 

Mayer et al. (2004) have therefore experimented with lowering
the coefficients of artificial viscosity, finding that for e.g.
$\alpha=0.5$ and $\beta=0$ or $\alpha=0.5$ and $\beta=1$ fragmentation
is more vigorous. However, the simulations of binary protoplanetary
disks in Mayer et al. (2005) were designed following a conservative
approach, hence the standard values $\alpha=1$ and $\beta=2$ were
employed. They do, however, include a modification due to Balsara
(1995), to modify the computation of the velocity divergence from its
usual form by multiplying the above equation by a correction factor

\begin{equation}\label{Balsara-f}
f_i = {{ | (\nabla \cdot {\bf v_i}) | } \over 
       { | (\nabla \cdot {\bf v_i}) | + | (\nabla \times {\bf v_i} ) |
               + 0.0001c_i/h_i}}.
\end{equation}

to reduce shear viscosity. This factor is near unity when the flow is
strongly compressive, but near zero in shear flows. In the simulations
of Nelson et al. (2000) and Nelson (2000), the typical reduction of
viscosity due to this term is a factor of three or better. 

Nelson's code starts from the same formulation of artificial viscosity
shown above (except that surface density replaces volume density),
modified by the same Balsara shear factor, but then also includes a
second treatment to obtain a locally varying artificial viscosity.
This treatment is due to Morris and Monaghan (1997), who implemented a
time dependence to the coefficient $\bar\alpha$ that allows growth in
regions where it is physically appropriate (strong compressions) and
decay in quiescent regions where it is inappropriate. The decay takes
place over distances of a few SPH smoothing lengths, after which the
coefficient stabilizes to a constant, quiescent value. Nelson adopts a
formulation including both the $\bar\alpha$ and $\beta$ terms, where
$\bar\alpha/\beta\equiv 0.5$, but the magnitudes of the coefficients
vary in time and space according to the Morris \& Monaghan
formulation. Thus, except in strongly compressing regions (shocks)
where it is required to stabilize the flow, artificial viscous
dissipation is minimized.

\begin{figure} 
\centering
\includegraphics[height=6cm]{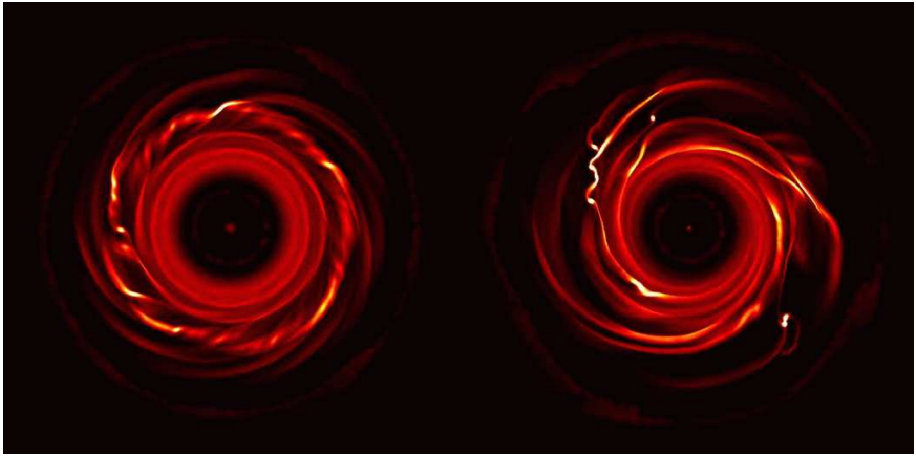}
\caption{Color coded density maps of two isothermal runs evolved with
1 million SPH particles. The disk starts with a minimum Toomre parameter
approaching $Q \sim 1$. The snapshots after about 8 orbital times at
$10$ AU, i.e. about 240 years, are shown. On the left a run without the
Balsara correction term is shown, on the right the same disk is evolved 
with the Balsara term  on.}
\label{fig:1}       
\end{figure}

In GASOLINE, only the Balsara correction term is added. Therefore, in
general the GASOLINE simulations presented in Mayer et al. (2005)
should be more diffusive, which should go in the direction of
suppressing fragmentation if all other aspects of the modeling are the
same. Figure 2 supports the claim of Mayer et al. (2004) on the effect
of artificial viscosity by showing that even the simple omission of
the Balsara term to reduce spurious shear viscosity can suppress
fragmentation in an otherwise fragmenting disk. However, we note that
the effect of artificial viscosity is more subtle than this, and that
tests such as those of Figure 2 probably address only the impact of
viscosity on the transition between mild overdensities and those
strong enough to begin collapsing. Once they start collapsing the
prevailing effect of artificial viscosity could actually be to enhance
the growth and survival of the clumps by taking kinetic energy out of
the system.

Artificial viscosity, be it explicitly inserted or implicit to the
code, is one way in which actual physical processes ocurring on the
sub-grid scale, such as shock front heating and dissipation, can be
included in the calculation. A tensor formulation artificial viscosity
is included in the Boss \& Myhill (1992) code, but is generally not
used in the disk instability models, except to discern the extent to
which artificial viscosity can suppress fragmentation (Boss 2006b).
Numerical stability is maintained instead by using as small a fraction
of the Courant time step as is needed, sometimes as low as 1\% of the
Courant value. Boss (2006b) showed that his code reproduced an
analytical (Burgers) shock wave solution nearly as well without any
artificial viscosity as when a standard amount ($C_Q = 1$, see Boss \&
Myhill 1992) of artificial viscosity was employed, implying that the
intrinsic numerical viscosity of his code was sufficient to stabilize
such a shock front. Tests on the full set of hydrodynamic equations
have not been published.

\subsection{Internal energy equation}

Both GASOLINE and Nelson's code employ the following energy equation
(called ``asymmetric''), advocated by Evrard (1990) and Benz (1990),
which conserves energy exactly in each pairwise exchange and is
dependent only on the local particle pressure,

\begin{eqnarray}
\frac{d\ u_i}{dt}& = & \frac{P_i}{\rho_i^2} \sum_{j=1}^{n}{m_j} 
\vec{v}_{ij} \cdot \nabla_i W_{ij}, 
\end{eqnarray}

\noindent
where $u_i$ is the internal energy of particle $i$, which is equal to
$1/(\gamma-1) P_i/\rho_i$ for an ideal gas. In this formulation
entropy is closely conserved making it similar to alternative entropy
integration approaches, such as that proposed by Springel \& Hernquist
(2002). The equation above includes only the part related to $PdV$
work, while the full energy equation has a term due to artificial
viscosity and one due to cooling to be discussed below.

The adiabatic index $\gamma$ is different in Nelson (2000) and Mayer
et al. (2005) because Nelson (2000) perform only two dimensional
simulations, In both cases the value of $\gamma$ appropriate for a gas
at temperatures $T < 1000$ K in which rotational transitions are
active, but not the vibrational ones, is assumed. Because of the
differences in dimensionality, this assumption yields $\gamma=1.4$ in
3D for the pure hydrogen gas (i.e. average molecular weight of 2.0)
assumed by Mayer et al. and $\gamma \sim 1.53$ in 2D, for the solar
composition gas (average molecular weight 2.31) used by Nelson (see
also Nelson 2006 for a discussion of issues that affect the ratio of
specific heats in 2D calculations). A value of $\gamma=1.42$ was
incorporated into the vertical structure models used in Nelson (2000)
in order to remain consistent, since the assumptions implicit in that
calculation required a 3D treatment. Other published work adopts
$\gamma=5/3$ (Rice et al. 2003a, 2005; Cai et al. 2006a,b). As we
explain below, the value of $\gamma$ can have an impact on whether
fragmentation occurs or not in a self-gravitating disk, both isolated
and in a binary system. A recent paper (Boley et al. 2006) explains
how using just one or two values of the adiabatic independently on the
actual temperature and density of the gas is probably a poor
approximation of the real physics of molecular hydrogen transitions
which can affect the outcome of gravitational instability. Future work
will need to asess that.

A term dependent on artificial viscosity is added to eq. (7)

\begin{equation}
\label{Piij}
\sum_{j=1}^{n}{m_j} {1 \over 2} \Pi_{ij}
\vec{v}_{ij} \cdot \nabla_i W_{ij}
\end{equation}

(see eq. (4) and (5) for the definition of $\Pi_{ij}$).

\noindent
This term allows the modeling of irreversible heating occurring in
shocks and the related changes in the entropy of the fluid. 

If no radiative cooling is included the resulting model is sometimes
dubbed ``adiabatic'' to distinguish it from the ``isentropic'' cases
in which no irreversible heating is included (Durisen et al. 2007).
However, in the simulations of binary protoplanetary disks of Mayer et
al. (2005) and Nelson (2000) a cooling term is always present. The
cooling term is described in the following section.

Boss \& Myhill (1992) code an equation for the specific internal
energy with explicit time differencing, in the same manner as the
other hydrodynamic equations are solved. This energy equation includes
the effects of compressional heating and cooling and of radiative
transfer, in either the diffusion or Eddington approximations. In the
latter case, a separate equation for the mean intensity must be
solved. In the diffusion approximation, the energy equation is solved
by Boss's code (Boss 2001) in the form

$${\partial (\rho E) \over \partial t} + \nabla \cdot (\rho E {\bf v}) =
- p \nabla \cdot {\bf v} + \nabla \cdot \bigl[ { 4 \over 3 \kappa \rho}
\nabla ( \sigma T^4 ) \bigr] $$

\noindent
where $\rho$ = gas density, $\bf v$ = fluid velocity, $E$ = specific
internal energy = $E(\rho, T)$, $p$ = gas pressure = $p(\rho, T)$,
$\kappa$ = Rosseland mean opacity of gas and dust = $\kappa(\rho, T)$,
$T$ = gas and dust temperature, and $\sigma$ = Stefan-Boltzmann
constant = $5.67 \times 10^{-5}$ cgs. The diffusion approximation
energy equation has been used for all of Boss's disk instability
models with radiative transfer to date. However, the Eddington
approximation energy equation was used to derive the initial
quasi-steady state thermal profiles (Boss 1996) used for the initial
conditions in Boss's disk instability models. In the Eddington
approximation code, the energy equation is

$${\partial (\rho E) \over \partial t} + \nabla \cdot (\rho E {\bf v}) =
- p \nabla \cdot {\bf v} + L $$

\noindent
where $L$ is the rate of change internal energy due to radiative
transfer. The formulation of $L$ depends on the optical depth $\tau$
as

$$L = 4 \pi \kappa \rho (J-B)  \ \ \ ......... \ \ \ \tau < \tau_c $$

$$L = {4 \pi \over 3} \nabla \cdot ({1 \over \kappa \rho} \nabla J)
\ \ \  ...... \ \ \ \tau > \tau_c  $$

\noindent
where $\tau_c$ is a critical value for the optical depth ($\tau_c \sim
1$), $\kappa$ is the mean opacity, $J$ is the mean intensity, and $B =
\sigma T^4 / \pi$ is the Planck function. The mean intensity $J$ is
determined by the equation

$$ {1 \over 3} {1 \over \kappa \rho} \nabla \cdot ( {1 \over \kappa \rho}
\nabla J) - J = -B $$

\noindent
The computational burden associated with the iterative solution of the
mean intensity equation in the Eddington approximation has so far
precluded its use in disk instability models with the high spatial
resolution needed to follow the evolution over many orbital periods.
Because of the high optical depths at the midplane (up to $\tau \sim
10^4$) of these disks, however, the diffusion approximation is valid
near the critical disk midplane, and radiative transfer in the
diffusion approximation imposes little added computational burden.

\subsection{Cooling in the simulations}
 
Nelson et al. (2000) and Nelson (2000) requires that the disk be in
instantaneous vertical entropy equilibrium and instantaneous vertical
thermal balance in order to determine its structure. This implicitly
assumes that the disk will be convectively unstable vertically over a
short timescale and quickly restore thermal balance. Convection is
expected given the high optical depths of massive gravitationally
unstable disks (Ruden \& Pollack 1996); vigorous vertical currents
with features resemblant of convective instabilities have been indeed
observed in massive protoplanetary disks modeled with either SPH and
grid codes (Boss 2003, 2004; Mayer et al. 2006). Under these
assumptions the gas is locally (and instantaneously) adiabatic as a
function of $z$. In an adiabatic medium, the gas pressure and density
are related by $p=K\rho^\gamma$ and the heat capacity of the gas,
$C_V$, is a constant (by extension, also the ratio of specific heats,
$\gamma$, see above). In fact, this will not be the case in general
because, in various temperature regimes, molecular hydrogen will have
active rotational or vibrational modes, it may dissociate into atomic
form or it may become ionized.

From the now known $(\rho, T)$ structure, Nelson et al. derive the
temperature of the disk photosphere by a numerical integration of the
optical depth, $\tau$, from $z=\infty$ to the altitude at which the
optical depth becomes $\tau=2/3$

\begin{equation}\label{tau-a}
\tau = 2/3 = \int_\infty^{z_{phot}}\rho(z)\kappa(\rho,T)dz.
\end{equation}
In optically thin regions, for which $\tau<2/3$ at the midplane, they
assume the photosphere temperature is that of the midplane. The
photosphere temperature is then tabulated as a function of the three
input variables radius, surface density and specific internal energy.
At each time the photosphere temperature is determined for each
particle from such table and used to cool the particle as a blackbody
at that temperature. The cooling of any particular particle proceeds
as

\begin{equation}
\label{dudt}
{{du_i}\over{dt}} = {{ -2\sigma_R T_{eff}^4 }\over{\Sigma_j} }
\end{equation}

where $\sigma_R$ is the Stefan-Boltzmann constant, $u_i$ and
$\Sigma_i$ are the specific internal energy and surface density of
particle $i$ and $T_{eff}$ is its photospheric temperature. The factor
of two accounts for the two surfaces of the disk. On every particle,
the condition that the temperature (both midplane and photosphere)
never falls below the 3~K cosmic background temperature is enfoced.
Rosseland mean opacities from tables of Pollack, McKay \&
Christofferson (1985) are used. Opacities for packets of matter above
the grain destruction temperature are taken from Alexander \& Ferguson
(1994).

In parts of the disk where the calculated midplane temperature is
greater than dust vaporization temperature, the opacity is temporarily
reduced to $\sim5$\% of its tabulated values over the entire column
above and below that point in the disk. This accounts for the fact
that dust formation, after once being vaporized, may occur at rates
slower than the timescales for vertical transport through the column.
In other regions of the disk we assume  that the opacity remains
unaffected.

Cooling is treated very differently in Mayer et al. (2005). Cooling is
independent on distance from the midplane also in this case (so there
is no dependence on $z$, as if there was constant thermal equilibrium
vertically) but there is an explicit dependence on the distance from
the center. The cooling term is proportional to the local orbital
time, $t_{orb} = 2\pi/\Omega$, where $\Omega$ is the angular velocity,
via the following equation

\begin{equation}
{\Lambda = dU/dt = U/A\Omega^{-1}}
\end{equation}

The disk orbital time is a natural characteristic timescale for spiral
modes developing in a rotating disk. Cooling is switched off inside 5
AU in order to maintain temperatures high enough to be comparable to
those in protosolar nebula models (e.g. Boss 1998), and in regions
reaching a density above $10^{-10}$ g/cm$^3$ to account for the local
high optical depth; indeed according to the simulations of Boss (2002)
with flux-limited diffusion the temperature of the gas evolves nearly
adiabatically above such densities. In practice in these regions the
gas simply obeys eq. (7) with the artificial viscosity term (8). Mayer
et al. (2005) considered cooling times going from $0.3$ to $1.5$ the
local orbital time. The jury is still out on whether the cooling times
adopted here are credible or excessively short. Recent calculations by
Boss (2002, 2004), Johnson \& Gammie (2003) and Mayer et al. (2006),
which use different approximate treatments of radiative transfer, do
find cooling times of this magnitude through a combination of
radiative losses and convection, but other works such as those of
Mejia et al. (2005), Cai et al. (2006) and Boley et al. (2006)
encounter longer cooling times and never find fragmentation. The
cooling times used in Nelson (2000) are about 25 times longer than the
typical orbital time in the region ($5-10$ AU), hence they were much
longer than in Mayer et al. (2005). As we shall see, this will have
profound implications on the final outcome of the simulations.

In work in progress, one of the authors, L.Mayer, has begun performing
simulations of binary protoplanetary disks using the new flux-limited
diffusion scheme for radiative transfer adopted in Mayer et al.
(2006), which adopts the flux-limiter of Bodenheimer et al. (1991) in
the transition between optically thick and optically thin regions of
the disk. The disk then radiates as a blackbody at the edge, with the
radiative efficiency being modulated by a parameter that defines how
large is the emitting surface area, or in other words, the part of the
disk that qualifies as edge. In section 5 we briefly describe some
preliminary results of these new calculations.

Boss (2001) noted that in his diffusion approximation models, the
radiative flux term is set equal to zero in regions where the optical
depth $\tau$ drops below 10, so that the diffusion approximation does
not affect the solution in low optical depth regions where it is not
valid. This assumption is intended to err on the conservative side of
limiting radiative cooling. A test model (Boss 2001) that varied this
assumption by using a critical $\tau_c = 1$ instead of 10 led to
essentially the same results as with $\tau_c = 10$, implying
insensitivity of the results to this assumption. Another test model
(Boss 2001) used a more detailed flux-limiter to ensure that the
radiative energy flux did not exceed the speed of light (specifically,
that the magnitude of the net flux vector $\vec H$ did not exceed the
mean intensity $J$), and also yielded essentially the same results as
the model with the standard assumptions. In low optical depth regions
such as the disk envelope, the gas and dust temperature is assumed to
be 50 K.

In the Boss (2006b) models, the disk was assumed to be embedded in an
infalling envelope of gas and dust that formed a thermal bath with a
temperature of 50K. Thus the effective surface boundary condition on
the disk is 50K. Boss (2004a) found that convective-like motions
occurred in the models with a vigor sufficient to transport the heat
produced at the midplane by clump formation to the surface of the
disk, where it is effectively assumed to be radiated away into the
protostellar envelope. This code also employs a full thermodynamical
description of the gas, including detailed equations of state for the
gas pressure, the specific internal energy, and the dust grain and
atomic opacities in the Rosseland mean approximation.

Molecular hydrogen dissociation is treated, as well as a parameterized
treatment of the transition between para- and ortho-hydrogen. Linear
interpolation from 100 K to 200K is used to represent the variations
between a specific internal energy per gram of hydrogen of $3/2 R_g T
/ \mu$ (where $R_g$ is the gas constant, $T$ is the temperature, and
$\mu = 2$ is the mean molecular weight for molecular hydrogen) for
temperatures less than 100 K and of $5/2 R_g T / \mu$ for temperatures
above 200 K. No discontinuities are present in this internal energy
equation of state, which has been used in all of Boss's disk
instability models.

Rafikov (2006) has noted that while vigorous convection is possible in
disks, the disk photosphere will limit the disk's radiative losses and
so may control the outcome of a disk instability. Numerical
experiments designed to further test the radiative transfer treatment
employed in the Boss models (beyond the tests described above) have
been completed, and other models are underway (Boss 2007, in
preparation). Understanding the extent to which the surface of a fully
3D disk, with optical depths that vary in all directions and with
corrugations that may shield the disk's surface from the central
protostar and other regions of the disk, requires a fully numerical
treatment and is not amenable to a simple analytical approach.

\section{Initial and Boundary Conditions}

Here we describe the initial and boundary conditions for the models, 
focusing on the different choices made by different workers. 
These choices include 
(1) density and temperature profiles of the disks, 
(2) disk masses and Toomre parameters,
(3) spatial resolution,
(4) the boundary conditions, and
(5) the orbital configuration in the binary experiments.

One important difference to point out is that Mayer et al. (2005)
start their binary simulations after growing the individual disks in
isolation. The disk mass is grown until it reaches the desired value
while keeping the temperature of the particles constant over time. The
Toomre parameter is prevented from falling below $2$ by setting the
temperature of the disk sufficiently high at the start. This way the
initial conditions used for the binary experiments are those of a
gravitationally stable disk. The Toomre parameter  is then lowered to
a value in the range $1.4-2$ by resetting the temperature of the
particles before placing the disk on the binary orbit (the temperature
is set to $65$ K, hence the Toomre parameter will depend on the mass
of the disk, see below in the next two sections).  As the disk evolves
in isolation, it expands slightly, losing the sharp outer edges. The
inner hole also fills up partially, but most of the particles remain
on very similar orbits. Once the disk is placed on an orbit around a
companion disk, the system is further evolved using the full energy
equation plus a cooling term dependent on the orbital time, therefore
including adiabatic compression and expansion, irreversible shock
heating and radiative cooling. Nelson (2000) do not grow the disk but
start with a treatment that also solves the full energy equation but
treats the cooling term differently (see above).

In both Mayer et al. (2005) and Nelson (2000) matter is set up on
initially circular orbits assuming rotational equilibrium in the disk.
The central star is modeled by a single massive, softened particle.
Radial velocities are set to zero. Gravitational and pressure forces
are balanced by centrifugal forces including the small contribution of
the disk mass.The magnitudes of the pressure and self-gravitational
forces are small compared to the stellar term, therefore the disk is
nearly Keplerian in character. No explicit initial perturbations are
assumed beyond computational roundoff error in either Mayer et al.
(2005) or Nelson (2000). Due to the discrete representation of the
fluid variables, this perturbation translates to a noise level of
order $\sim 10^{-3}-10^{-2}$ for the SPH calculations. The relatively
large amplitude of the noise originates from the fact that the
hydrodynamic quantities are smoothed using a comparatively small
number of neighbors (see Herant \& Woosley 1994). An increase in the
number of particles does not necessarily decrease the noise unless the
smoothing extends over a larger number of neighbors. This perturbation
provides the initial seed that can be amplified by gravitational
instability. The initial disk model in Boss (2006b) is rotating with a
near-Keplerian angular velocity chosen to maximize the stability of
the initial equilibrium state, with zero translational motions. The
envelope above the disk, however, is assumed to be falling down with
free-fall velocities but with the same angular velocity profile as the
rotating disk. Random cell-to-cell noise at the level of 10\% is
introduced to the disk density to seed the cloud with non-axisymmetry
at a controlled level.

\subsection{Density and temperature profiles}

Mayer et al. (2005) grow the disks slowly over time until the desired
mass is reached. The initial disk models extend from 4 to 20 AU and
have a surface density profile $\Sigma(r) \sim r^{-1.5}$ with an
exponential cut-off at both the inner and outer edge. The initial
vertical density structure of the disks is imposed by assuming
hydrostatic equilibrium for an assumed temperature profile $T(r)$.
Nelson (2000) adopts a power law for the initial disk surface density
profile of the form 

\begin{equation}\label{cooldenslaw}
\Sigma(r) = \Sigma_0 \left[ 1 + \left({r\over r_c}\right)^2\right]^
{-{p\over{2}}},
\end{equation}

\noindent
where $p$ is $3/2$ and $\Sigma_0$ is the central surface density of
the disk, which is determined once the total disk mass is assigned.
The Nelson (2000) disk extends from $0.3$ to $15$ AU, and the core
radius used for the power laws to the value $r_c=1$~AU. The stars had
Plummer softening of $0.2$ AU each, while a softening length of 2~AU
was used with the spline kernel softening in Mayer et al. (2005).

The shape of the initial temperature profile in Mayer et al. (2005) is
similar to that used by Boss (1998, 2001) and is shown in Figure 3
together with the profiles of several runs with either binary or
isolated disks after a few orbits of evolution.

\begin{figure}
\centering 
\includegraphics[height=7cm]{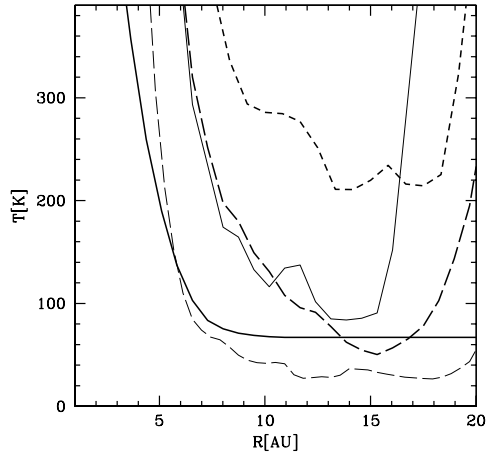}
\caption{Azimuthally averaged mid-plane temperature profiles at the
time of maximum amplitude of the overdensities (at between 120 years
and 2000 years of evolution depending on the model) in some of the
runs described in Mayer et al. (2005). The initial temperature profile
is shown by the thick solid line. We show the results for a run with
two massive disks ($M=0.1 M_{\odot}$, thick short-dashed line, the
disks do not fragment) at a separation of about $60$ AU, a run with
just one of these two disks in isolation (thin long-dashed line, disk
fragments), a run with the same massive disks at a larger separation
of $116$ AU (disks fragment) and a run with two light disks ($M=0.01
M_{\odot}$, the disks do not fragment) at a separation of $60$ AU. The
runs adopt cooling times in the range $0.5-1.5$ the local orbital
time. A smaller separation and a larger disk mass both favour stronger
spiral shocks and hence a larger temperature increase during the
evolution.} \label{fig:1}       
\end{figure}

The temperature depends only on radius, so there is no difference
between midplane and an atmosphere. Between 5 and 10 AU the
temperature varies as $\sim r^{-1/2}$, which resembles the slope
obtained if viscous accretion onto the central star is the key driver
of disk evolution (Boss 1993). Between 4 and 5 AU the temperature
profile rises more steeply, in agreement with the 2D radiative
transfer calculations of Boss (1996), while it smoothly flattens out
for $R > 10$ AU and reaches a constant minimum temperature (an
exponential cut-off is used). The initial temperature profile of
Nelson (2000) is instead

\begin{equation}\label{cooltemplaw}
T(r) = T_0\left[1 + \left({r\over r_c}\right)^2\right]^{-{q\over{2}}},
\end{equation}

\noindent
where $q$ is $1/2$ and T$_0$ is the central temperature, which is
determined once the minimum desired Toomre parameter, and hence the
minimum temperature, is determined (see above). 

In Mayer et al. (2005) the minimum temperature is fixed to $65$ K. It
is implicitly assumed that the disk temperature is related to the
temperature of the embedding molecular cloud core from which the disk
would be accreting material (Boss 1996). Note that, at least for the
protosolar nebula, $50$ K is probably a conservative upper limit for
the characteristic temperature at $R > 10$ AU based on the chemical
composition of comets in the Solar System (temperatures as low as $20$
K are suggested in the recent study by Kawakita et al. 2001). Outer
temperatures between $30$ and $70$ K are found also for several T
Tauri disks by modeling their spectral energy distribution assuming a
mixture of gas and dust and including radiative transfer (D'Alessio et
al. 2001).

In the Boss (2006b) models, the initial disk is an approximate
semi-analytic equilibrium model (Boss 1993) with a temperature profile
derived from the Eddingtion approximation radiative transfer models of
Boss (1996). The outer disk temperatures were taken to be either 40K,
50K, 60K, 70K, or 80K, in order to test the effects of starting with
disks that were either marginally gravitationally unstable (40K, 50K)
or gravitationally stable (60K, 70K, 80K). The disk temperature is not
allowed to fall below its initial value, an approximation that errs on
the side of suppressing fragmentation. 

Figure 4 shows the initial radial temperature profile for the models
with an outer disk temperature of 80 K, as was assumed in the Boss
(2006b) model shown in Figure 7. The temperature rises strongly toward
the protostar at 0 AU because of the heating associated with mass
accretion onto the disk from the protostellar envelope and onto the
central protostar from the disk (Boss 1993, 1996). The temperature
distributions in Boss (1993, 1996) are steady state solutions for
axisymmetric (2D) protoplanetary disks with varied disk and stellar
masses, opacities, and other parameters.

\begin{figure}
\centering 
\includegraphics[height=10cm]{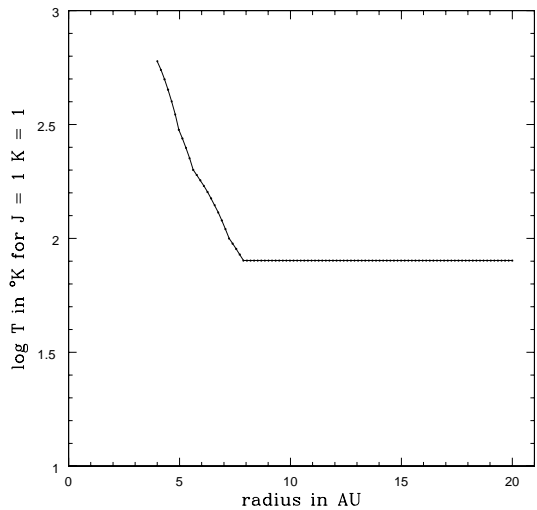}
\caption {Initial radial temperature profile for the midplane of the
Boss (2006b) models with outer disk temperatures of 80K. Each dot
corresponds to a radial grid point (100 in all) distributed between 4
AU and 20 AU. This initial profile was used for the model shown in
Figure 7.} \label{fig:1}       
\end{figure}

\subsection{Disk masses and Toomre parameters}

Both Nelson (2000) and Mayer et al. (2005) consider a system composed
of two disks with their central stars, while Boss (2006) models one
system with a disk and star, and the companion with just a star. The
two disks have equal masses, $0.05 M_{\odot}$, in Nelson (2000), while
Mayer et al. (2005) consider a range of disk masses encompassing
models from as light as the lowest expected values of the minimum mass
solar nebula ($0.012 M_{\odot}$) to as massive as the heaviest among T
Tauri disks ($0.1 M_{\odot}$. A few simulations with disks having
unequal masses were performed by Mayer et al. (2005), while in the
majority of the runs the disks have the same mass. Boss (2006)
considers systems having equal mass stars. The mass of the central
star is $0.1 M_{\odot}$ in Mayer et al. (2005) and $0.5 M_{\odot}$ in
Nelson (2000). The minimum Toomre parameter is $\sim 1.5$ in Nelson
(2000) achieved just inside the outer edge of the disk, at $10-12$ AU,
while it is  $Q_{min} \sim 1.4$ or higher in Mayer et al. (2005) at
the disk edge, where the temperature also falls to its minimum. The
details of the individual models can be found in Mayer et al. (2005).
The initial surface densities and Toomre profiles differ between the
two works and give rise to a different susceptibility to various
channels by which non-axisymmetric models can grow. In Nelson (2000),
$Q$ is nearly flat over the largest portion of the disk, with a steep
rise at small radii and a shallow increase towards the outer edge of
the disk. In Mayer et al. (2005), disks are constructed in such a way
that they begin with a steep inner rise of $Q$, which decreases
outwards and reaches its minimum at the disk edge (Mayer et al. 2004).
However, as the disk grows in mass the $Q$ profile changes; its
minimum shifts inwards, near $15$ AU, so that the overall profile
becomes quite similar to that of Nelson (2000) by the beginning of the
simulations. We refer to Mayer et al. (2004) and Nelson et al. (2000)
for details on the $Q$ profiles. 

In the Boss (2006b) models, the disk mass is $0.091 M_\odot$ and the
central protostar has a mass of $1 M_\odot$. Because the outer disk
temperatures in different models are varied from 40K to 80K, the
minimum value of the Toomre $Q$ parameter varies from 1.3 to 1.9 in
the models.

Figure 5 shows the radial profile of the Toomre Q parameter for the
Boss (2006b) models with an outer disk temperature of 80K, the same as
for the model shown in Figures 4 and Figure 7. The disk is very stable
to gravitational perturbations in its inner regions, because of the
high inner disk temperatures (Figure 4), but drops to a minimum value
of 1.9 in the outer disk.

\begin{figure}
\centering 
\includegraphics[height=10cm]{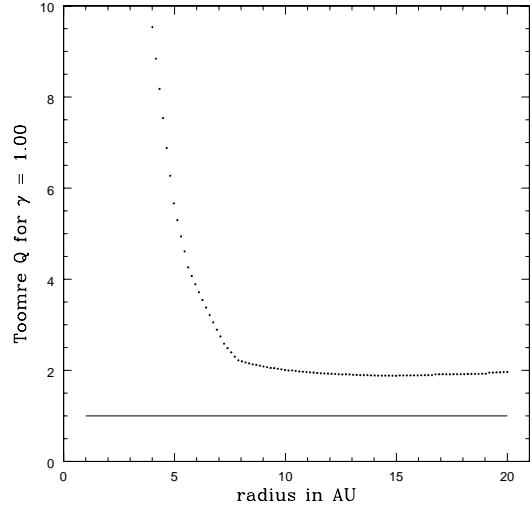}
\caption { Initial Toomre Q profile for the midplane of the Boss
(2006b) models with outer disk temperatures of 80K, used for the
model shown in Figure 7}
\label{fig:1}       
\end{figure}

\subsection{Numerical resolution}

Nelson (2000) employs 60000 particles per disk in his 2D simulations,
while Mayer et al. (2005) used 200000 particles per disk. Due to the
differences in dimensionality, the mass resolution along the disk
midplane is quite similar, so in this respect the two simulations are
quite comparable. The gravitational softening is fixed at $\sim 0.06
AU$ in Mayer et al. (2005), while it can become smaller than that in
Nelson (2000) in overdense regions forming during the simulation.

Boss (2006b) used a grid with either $100 \times 22 \times 256 = 0.56
\times 10^6$ or with $100 \times 22 \times 512 = 1.1 \times 10^6$
grid points, distributed over the top hemisphere of the calculational
grid, and either 32 or 48 terms in the spherical harmonic expansion
for the gravitational potential.

\subsection{Boundary conditions}

Mayer et al. (2005) adopt no boundary conditions at all in their
disks. The central star is free to move in response to deviations of
the gravitational potential of the disk from the initial equilibrium
and gas particles can get as close as resolution allows. Although such
a choice is not ideal from the point of view of computational
efficiency, since particles nearest to the center have the shortest
timesteps, this should reduce fluctuations in the inner density and
pressure profiles due to the sudden removal of particles. Nelson
(2000) instead implements an inner boundary condition by treating the
central star as a sink particle, namely a particle that absorbs the
mass and momentum of particles falling below some threshold radius. He
uses an accretion radius of $0.2$ AU as a compromise between the
numerical requirement that the integration time step not be so small
that long period evolution cannot be followed and the desire to model
as large a radial extent of the disk as possible. The gravitational
softening of the central star in Mayer et al. (2005) is  $2$ AU, which
for a spline kernel softening means that effectively the force
resolution is $4$ AU. In both Mayer et al. (2005) and Nelson (2000)
the initial location of the innermost ring of particles lies slightly
outside the inner accretion radius or gravitational softening of the
star. There is no outer boundary condition in either of these works. 

In the Boss (2006b) models, the inner boundary at 4 AU is allowed to
remove mass and angular momentum from the grid and deposit it onto the
central protostar. The outer boundary at 20 AU is fixed in space, and
attempts to capture gas which reaches it while suppressing its
tendency to bounce back inward. As a result of the strong tidal forces
by the binary companions in Boss (2006b), disk gas which attempts to
flow outward becomes artificially trapped in the outermost shell of
cells. Thus any clumps observed on the or close to the outer boundary
of the Boss (2006b) are artifacts of the outer boundary conditions and
should disregarded.

\subsection{Orbital parameters}

Both Mayer et al. (2005) and Nelson (2000) consider coplanar disks
corotating with their orbital motion as expected from fragmentation of
a cloud core (Bate 2000). If core formation and fragmentation is a
highly dynamical process as recent simulations of gravoturbulent
molecular cloud collapse suggest, so that several cores interact
strongly during collapse, more complicated orbital configurations will
arise. These will need to be explored in the future. Fast close
encounters between disks will also occur in the latter scenario, as
Lodato et al. (2007) have investigated. In Nelson (2000) the
semi-major axis is set to an initial value of $50$ AU and the orbit
has an eccentricity of $0.3$. Mayer et al. (2005) consider a more
circular orbit ($e=0.14$) and two different semi-major axes, of $58$
and $116$ AU. The calculation starts with the companion disk being at
the apoastron of the orbit.

Boss (2006b) considers only a single disk in his models, with the
binary companion being a point mass protostar. The binary companion
orbits were chosen to have semimajor axes of either 50 or 100 AU,
namely comparable to those chosen by Mayer et al. (2005), to have
eccentricities of either 0.25 or 0.5, and to having the calculation
start off with the binary companion at either apoastron or at
periastron in its orbit. The models assume that the mass of the binary
companion is the same as that of the central protostar, $1 M_\odot$.

Note that in the Boss (2006b) models, the binary semimajor axis is
defined to be equal to the radial separation between the two
protostars, so that in a model with a semimajor axis of 50 AU and an
eccentricity of 0.5, the closest approach between the two protostars
is 25 AU, just beyond the edge of the 20 AU-radius disk that is being
perturbed by the binary companion. If the binary companion had a
similar size disk, these disks would collide, but the binary companion
is assumed to be a diskless point mass in all of the Boss (2006b)
models.

\section{Gravitational instability in binary systems}

\subsection{Does binarity help or suppress disk fragmentation?}

The three studies seemingly reached very different conclusions
regarding the role of binary companions in disk instabilities. Nelson
(2000) found no aid to fragmentation, Mayer et al. (2005) found
fragmentation in a few cases but also found an indication that
binarity might reduce the susceptibility to fragment, while Boss
(2006b) found fragmentation to be enhanced by binarity. However, we
shall analyze in more detail these differences, trying to understand
their causes.

As we have seen in the previous sections, there are many differences
in the codes and setup of the numerical experiments. Mayer et al.
(2005) performed a larger number of experiments, thus exploring a
larger parameter space in terms of initial conditions. Nelson (2000)
and Boss (2006b) had more realistic treatments of radiation transfer,
while Boss (2006b) ran the highest resolution experiments. Disk
thermodynamics is crucial for the outcome of gravitational
instability. Fragmentation will occur only if cooling times are
comparable to the orbital time. Therefore, leaving alone all the other
differences, the simple fact that Nelson (2000) had cooling times in a
large fraction of his disks that were a factor of 10 times longer than
the orbital times can explain why fragmentation did not occur in his
models. With such long cooling times, disks will not fragment in
isolation, no matter how strong is the spiral structure appearing in
the disk, and thus irrespective of whether this structure is
spontaneous or is tidally triggered by a nearby companion.
Nevertheless, fragmentation is not determined by the longest cooling
times in the disk, but by the shortest. Nelson's disks did exhibit
short cooling times at larger radii, in spite of this fact, did not
fragment.

Rapid cooling of the disk midplane by convective-like motions in 3D
disks has been shown to occur with several different codes (Boss
2004a, 2005; Boley et al. 2006; Mayer et al. 2006) and can lead to
disk fragmentation (Boss 2004a, 2005), provided that the heat
transported upward by these motions to the disk's surface can be
radiated away to the protostellar envelope, a condition disputed by
Rafikov (2006). The vertical structure model of Nelson (2000) assumed
efficient vertical energy transport via convection, but did not
produce fragmentation.  The thermal boundary conditions on the disk
surface then become of critical importance, and these boundary
conditions are the subject of current research. Until the issue of
disk thermal boundary conditions can be further clarified, it is
useful to ask whether for relatively short cooling times, comparable
to or less than the orbital period, binarity promotes or suppresses
fragmentation. The latter question is what both Mayer et al. (2005)
and Boss (2006b) tried to answer. 

\begin{figure}
\centering 
\includegraphics[height=4.5cm]{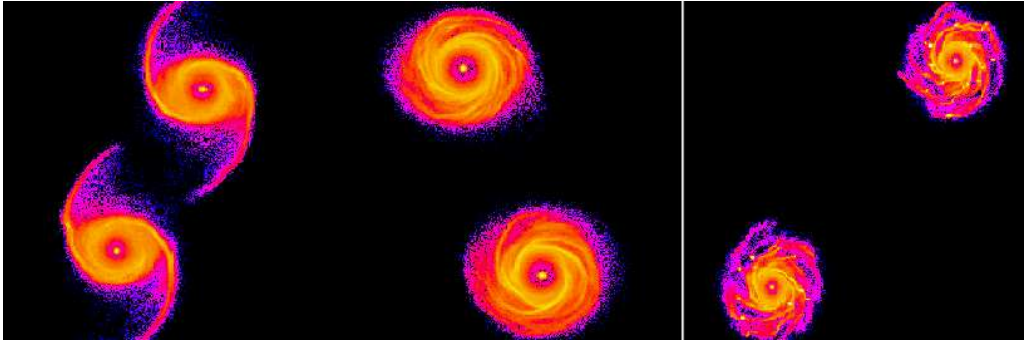}
\caption{Face-on density maps for two simulations of interacting
$M=0.1 M_{\odot}$ protoplanetary disks in binaries with $t_{cool} =
0.5P_{rot}$ viewed face-on. The binary in the left panel has a nearly
circular binary orbit with an initial separation of 60 AU and is shown
after first pericentric passage at 150 yrs (left) and then at 450 yrs
(right). Large tidally induced spiral arms are visible at 150 yrs. The
right panel shows a snapshot at 160 yrs from a simulation starting
from an initial orbital separation that is twice as large. In this
case, fragmentation into permanent clumps occurs after a few disk
orbital times. Figures adapted from {\it Mayer et al.} (2005). }
\label{fig:1}       
\end{figure}

Mayer et al. (2005) explored a range of cooling times, from less than
0.5 to 1.5 times the orbital period. They  found that the effect of
binarity changed with disk mass: except for the shortest cooling time
($0.3 T_{orb}$), massive disks $M_{disk} = 0.1 M_{\odot}$ that
fragmented in isolation did not fragment when in a binary with a
separation of $\sim 60$ AU, while disks with masses $0.05-0.08
M_\odot$ that do not fragment in isolation do fragment in a binary
system with a separation of $\sim 60$ AU provided that the cooing time
is somewhat shorter than the disk orbital time. When the separation
grows from $60$ to  $116$ AU the behaviour of the disks becomes almost
indistinguishable from that seen in isolation, and now fragmentation
becomes possible in the $0.1 M_{\odot}$ disks. Simulations from this
work are presented in Figure 6, which shows how larger separations are
more favourable to fragmentation in the case of massive disks. Mayer
et al. interpreted the different behvaiour of disks having different
masses  as the product of different net cooling times at different
mass scales. In more massive disks the spiral arms grow stronger as
they are better amplified by self-gravity; as a result, shocks are
more oblique and disk material acquires higher eccentricities,
resulting in overall higher Mach numbers and stronger heating. For a
given cooling time, the ``net cooling'', namely the ratio between
cooling and heating, is higher for lighter disks. Below some mass,
however, the self-gravity is so low that spiral arms cannot grow
enough to trigger fragmentation, no matter how strong is the
perturbation of the companion and even if the cooling time is
comparable to or shorter than the disk orbital time. Figure 3 shows
the temperature evolution of the disk in some of the runs performed in
Mayer et al. (2005). It shows that temperature increase in the outer
disk, which opposes fragmentation, is larger in more massive disks and
at smaller binary separation, supporting the interpetration of the
authors concerning why fragmentation can be suppressed. The results of
Mayer et al. (2005) are not in conflict with those of Nelson (2000)
for runs that have similar orbital parameters and comparable disk
masses, i.e. disk/star systems with mass ratios of $0.1$ and
separations of $50-60$ AU. It is true that in some of the latter runs
disks fragment in Mayer et al. (2005) while they never fragment in
Nelson (2000), but this discrepancy is seen only for the shortest
cooling times ($0.3-0.5 T_{orb}$) used in Mayer et al. (2005), these
being more than an order of magnitude shorter than the typical cooling
times of Nelson (2000). 

Boss (2006b) used his standard radiative transfer approach to handle
the disk thermodynamics, i.e., diffusion approximation radiation
transport, Rosseland mean dust opacities, and detailed equations of
state for the gas pressure and specific internal energy. One model
from Boss (2006b) was particularly similar to that of Nelson (2000):
the binary companions had semimajor axes of 50 AU in both models, and
an eccentricity of 0.3 in Nelson (2000) and 0.25 in Boss (2006b). The
initial value of Toomre's $Q$ was $\sim 2$ throughout most of the
inner disk in Boss (2006b) and in Nelson (2000), implying that both
disks were initially stable. While the total system mass was twice as
high in Boss (2006b) as in Nelson (2000), the ratio of the disk mass
to the protostar mass was the same in both models, $\sim$ 0.1. In the
case of Nelson (2000), the disk formed strong spiral arms but never
fragmented. It heated up as a result of of viscous dissipation and
partially also because of the spiral shocks, finally reaching a steady
state characterized by a Toomre Q in the range of 4 to 5, i.e., quite
stable to the growth of gravitational perturbations. (Figure 8). By
comparison, in the Boss (2006b) model (see Figure 7)

\begin{figure}
\centering 
\includegraphics[height=10cm]{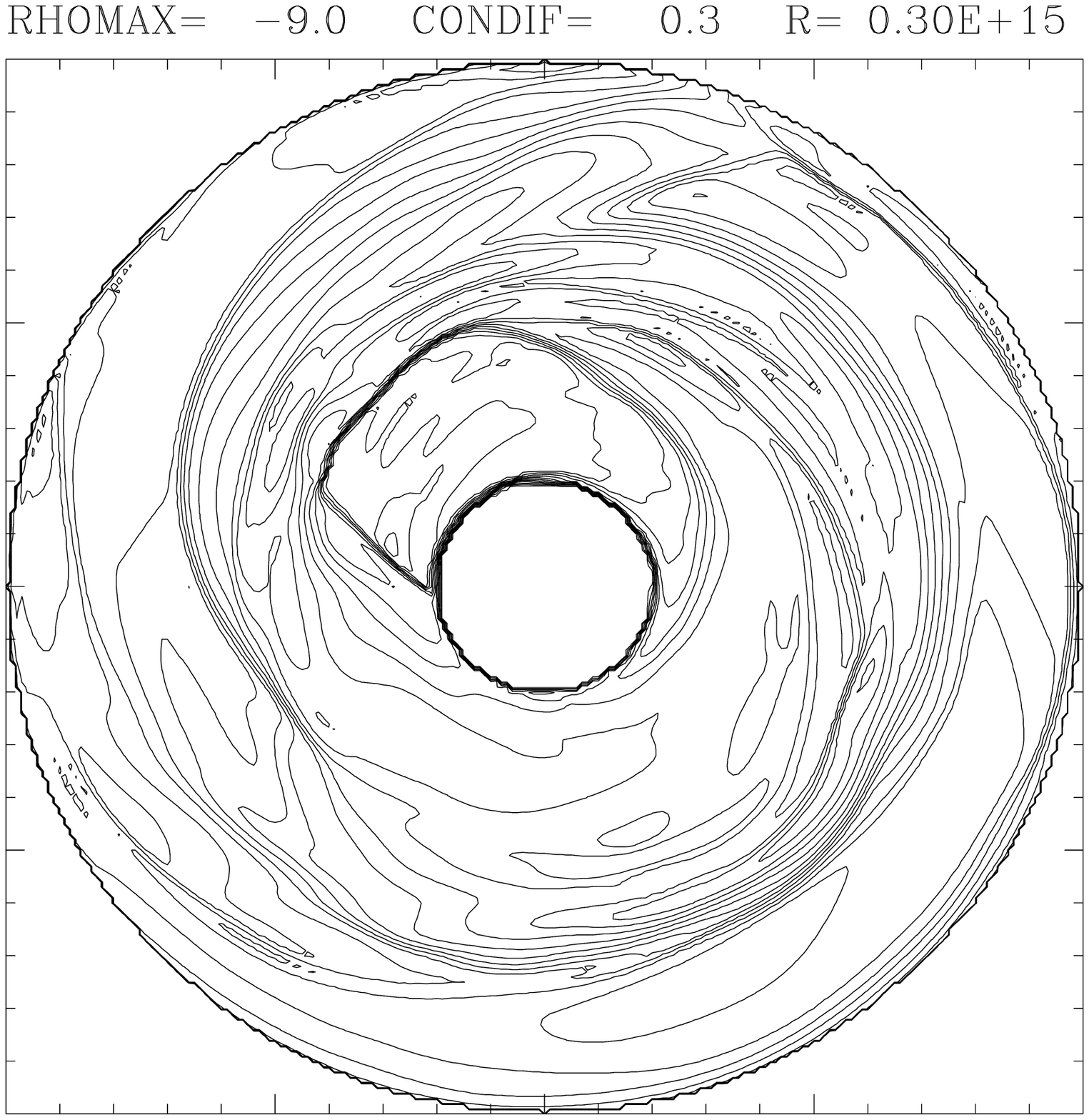}
\caption{Midplane density contours after 241 yr of evolution of a
$0.091 M_\odot$ disk in orbit around a member of a binary $1 M_\odot$
protostar system, showing the formation of a self-gravitating clump of
mass $4.7 M_{Jup}$ at 10 o'clock (Boss 2006b)} 
\label{fig:1}       
\end{figure}

the disk also formed strong spiral arms, but a self-gravitating clumps
was able to form as well, with a mass of 4.7 $M_{Jup}$. When Boss
(2006b) ran an identical model, except with the binary companion on a
more eccentric orbit ($e = 0.5$), the clump that formed at a similar
time to the one in Figure 5 (above) was not massive enough (only 0.68
$M_{Jup}$) to be self-gravitating, though later in the evolution
self-gravitating clumps did form. These two models show that in the
models of Boss (2006b), the ability of a binary companion to induce
disk fragmentation depends strongly on the orbital parameters of the
companion.

\begin{figure}
\centering 
\includegraphics[height=7cm]{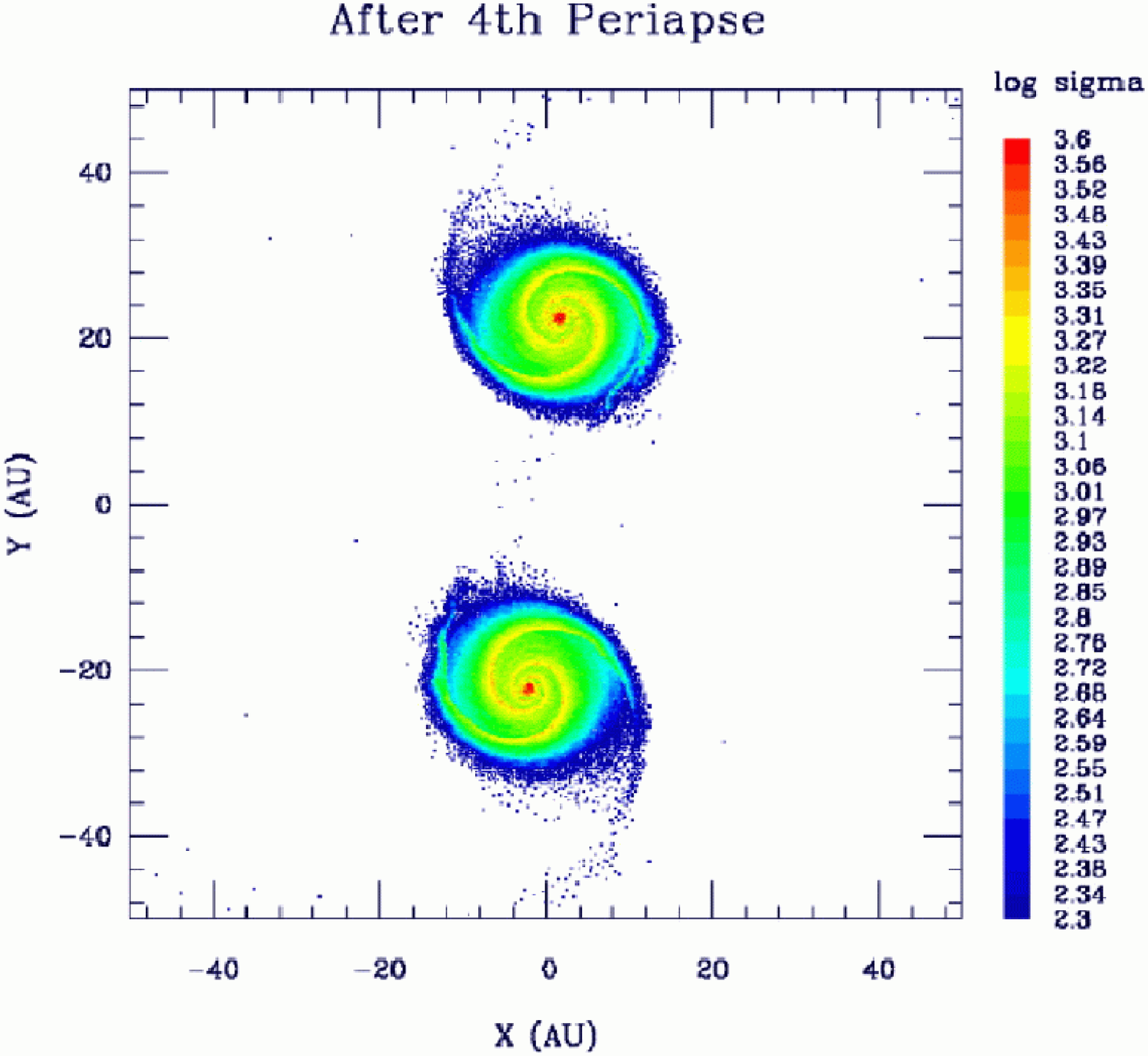}
\caption{Particle distribution of the binary system after periapse
passage. Mass surface density units are in log (g cm$^{-2}$). The
trajectory of each component is counterclockwise, and periapse occurs
when the stars (at each disk center) reach the $y=0$ axis and are $35$
AU apart. The tidal torques have caused two-armed spiral structures to
develop in the disks and significant mass redistribution, but no
fragmentation. The figure has been adapted from Nelson (2000). }
\label{fig:1}       
\end{figure}

It is important to note, however, that while Nelson (2000) was able to
follow the evolution of the disks through several periastron passages,
and monitor the resultant disk heating, Boss (2006b) only followed a
single periastron, largely because of the pile-up of disk mass at the
outer edge of the disk, an obvious numerical artifact that greatly
reduced the value of carrying the models any further in time. Thus it
is uncertain what would happen to the clumps that formed after the
first periastron passage in the Boss (2006b) models, if subsequent
periastron passages were calculated as well.

The models of Boss (2006b) investigated binary companions having
eccentrcic orbits with semimajor axes of 50 or 100 AU, while those of
Mayer et al. (2005) explored primarily $58$ AU (13 models), as well as
4 models with $a = 116$ AU. The models with $a=115$ AU fragment
similarly to the $100$ AU models in Boss (2006b). On the other end,
for the models with $a=58$ AU Mayer et al. (2005) found that whether
or not fragmentation occurred depended on the disk masses and assumed
cooling times, as described previously. The models with $a=58$ AU
separation, disk masses of 0.1 $M_\odot$ and protostars with masses of
1 $M_\odot$, can be compared directly with some of Boss's models
having essentially the same parameters. While such models never
fragment in Mayer et al. (2005) or produce transient clumps that
disappear in a few disk orbital times, fragmentation always occurs in
Boss (2006b). One subtletly in the comparison is that indeed even all
the fragments obtained by Boss (2006b) are also transient fragments,
because his finite-difference code is unable to provide the enhanced
local spatial resolution that is needed to allow self-gravitating
clumps to survive and orbit indefinitely. On the other end, Mayer et
al. (2005) did not run the same model at higher resolution as they had
done for isolated disk models in previous works (e.g. Mayer et al.
2004), and therefore one cannot exclude that their clumps would
survive longer or fragmentation would be aided in the first place with
an increased number of particles and proportionally smaller softening
length. The azimuthal resolution in Boss (2006b) is indeed higher than
that the hydrodynamical resolution and, even more, than the
gravitational force resolution adopted by Mayer et al. (2005). In
Mayer et al. (2005) mild overdensities build up along the spiral arms
after periastron  even in the runs that do not fragment (Figure 6),
but are immediately dissolved; it is possible that with higher
resolution they would become more nonlinear and collapse based on the
preliminary results of the aforementioned code comparison between AMR
and SPH codes (Mayer et al., in preparation). Nevertheless, the
difference remains that a companion on a tighter orbit suppresses
fragmentation according to Mayer et al. (2005), while it promotes it
according to Boss (2006b). The intense heating is the reason of such
suppression of fragmentation in Mayer et al. (2005), and such heating
is apparently not present in Boss (2006b). Whether the SPH artificial
viscosity is biasing the results too much with the stronger shocks
present in binary systems or whether the disks in Boss (2006b) cool
too fast is at the moment unclear. It is especially noteworthy that
Nelson (2000) studied a 50~AU binary system with SPH, at lower
resolution than in the Mayer et al. work. This system should therefore
be even more strongly affected by the presence of heating from
artificial viscosity. In spite of this, Nelson finds that in fact the
disks are {\it too cold}, compared to the observed L1551~IRS5
system--implying still more heating is required than artificial
viscosity provides, making his disks even less likely to fragment. 

Another difference in the setup that might explain the discrepancies
is the fact that in Boss (2006b) the companion is simply a protostar,
while it is a disk with a protostar in Mayer et al. (2005). In tighter
binary systems the presence of the other disk might have an effect;
indeed near periastron the two disks almost touch each other, possibly
enhancing tidal and compressional heating on one another compared to
the case in which only a companion star is present. Since the orbits
in Mayer et al. (2005) are nearly circular, in the $58$ AU case the
two disks are almost always in the latter situation. This would also
mean, however, that the behavior of real tight binary systems, which
will normally have two disks orbiting each other, should  be closer to
what found by Mayer et al. (2005) and Nelson (2000).

\subsection{Disk evolution: internal vs. external}

The question also arises of how much of the disk restructuring is due
to the disk's self-gravity and how much to tidal torques induced by
the companion. Figure 9 shows the evolution of the disk surface
density profile in one of the binary disks simulations of Mayer et al.
(2005) evolved both with and without self-gravity. Clearly some mass
transport has happened even without self-gravity. Such mass transport
is the result of tidal torques induced by the gravitational
interaction with the companion and the fact that little disk material
exists in the initial state inside 4~AU to resist the flow of
additional material inward. These tidal torques produce a two-armed
spiral mode in the otherwise passive disk. The disk becomes truncated
to a smaller radius and more mass piles up in the inner few AU as the
arms redistribute angular momentum. We recall that the disks is Mayer
et al. (2005) have been slowly grown in mass in isolation before being
evolved with a companion; while the disk evolves in isolation the
inner hole present in the initial conditions is gradually filled, and
therefore the rapid accumulation of mass seen in the binary case is
not an artificial result of the inner boundary condition. The mass
inflow produces compressional heating, raising the temperature of the
disk inside $10$ AU. Exchange of mass between the two disks occurs but
their mass varies by only $\sim 10\%$. 

Despite the fact that the tidal interaction modifies the disk
structure irrespective of disk self-gravity, Figure 9 shows that such
changes are moderate compared to those occurring when self-gravity is
included. When self-gravity is included the density peak that develops
is almost a factor of 2-3 higher than the maximum density in similar
disk models evolved in isolation (Mayer et al. 2004). This statement
applies to all of the runs in Mayer et al. (2005).

\begin{figure}
\centering 
\includegraphics[height=7cm]{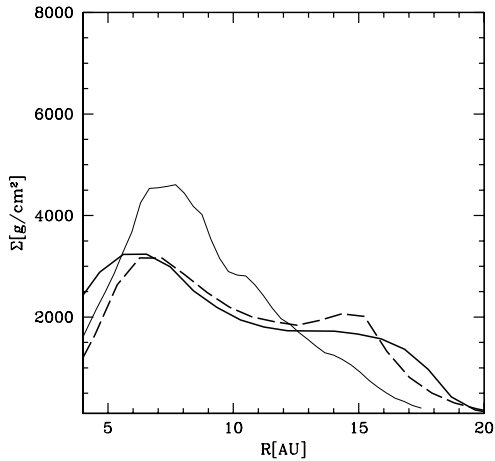}
\caption{Azimuthally averaged surface density profile of a disc with
mass $M_d = 0.01 M_{\odot}$ at $t=0$ (solid line) and for its evolved
state after being run without self-gravity for two orbits around an
equally massive companion (dashed line). The surface density of a run
employing a massive disk ($M=0.1 M_{\odot}$) evolved with self-gravity
is shown instead by the thin solid line. A cooling time equivalent to
$0.3$ the orbital time was adopted in both runs. The figure has been
adapted from Mayer et al. (2005).} \label{fig:1}       
\end{figure}

The larger density may explain why models with masses lower than $0.1
M_{\odot}$ become more prone to fragmentation when perturbed by a
binary companion; evidently in these lighter disks the heating from
shocks is not enough to compensate such a large density increase.
Since disks are truncated within 15 AU, when clumps form they do so
within such radius, typically between 8 and 12 AU. The locations where
they form correspond to the location of the density maximum and are
slightly closer to the star compared to those of clumps in the
isolated disks studied by Mayer et al. (2004). In fact in isolation
gravitationally unstable disks typically develop a density maximum
between 12 and 15 AU, and that is where $Q$ drops below 1 and
fragmentation  occurs (Mayer et al. 2004). The conclusion is that in
all the simulations of Mayer et al. (2005) the restructuring of the
disk results from a combination of tidal torques and  self-gravity of
the disks.

\subsection{Temperatures in binary self-gravitating disks 
and effects on dust grains}

Both Nelson (2000) and Mayer et al. (2005) found significant heating
along spiral shocks in binary systems. Mayer et al. (2005) found that
the temperatures can be a factor of 2-3 higher relative to the same
disk in isolation for disks in the mass range $M_{disk} \sim 0.05-0.1
M_{\odot}$. This has important implications for the destruction of
dust grains, hence on the formation of planetesimals, and thus of
those Earth-sized rocky cores that are a necessary step to form giant
planets in the core-accretion model. The consequence of the high
temperatures in the GI active outer region of the disk is the
vaporization  of ice grains, which constitute as much as $30-40\%$ of
the dust content in the disk. The actual surface density of solids
might be reduced by up to $40 \%$ compared to isolated disks. The
direct consequence should be that core accretion will be less
efficient in binary systems compared to isolated disks. The results of
Nelson (2000) are shown in Figure 10. Strong heating is instead absent
in light disks ($M_{disk} 0.01 M_{\odot}$, whose temperatures increase
by less than $50\%$. Therefore core accretion should be favored in
such light disks, with masses comparable to the minimum mass solar
nebula, because a larger relative dust content should be maintained.
Those, however, are also the disks in which the growth of  rocky cores
of a few Earth masses (necessary to trigger runaway gas accretion)
would take longer owing to their low surface densities.

\begin{figure}
\centering 
\includegraphics[height=7cm]{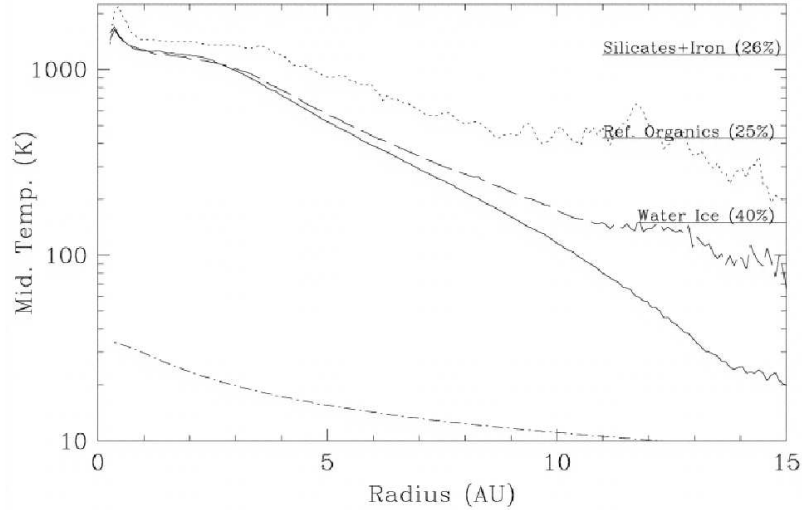}
\caption {Temperature profiles of the disks in Nelson (2000) shown
before (solid line) and after (dashed line) the 4th periapse. The
initial profile is shown with a dashed-dotted line. The dotted line
shows the maximum temperature reached inside the spiral arms at that
radius. At the right are vaporization temperatures of the major grain
species in the solar nebule and their fraction of the total grain mas,
as discussed in Pollack et al. (1994). } 
\label{fig:1}       
\end{figure}

A caveat in the above discussion is that the surface density of
massive disks ($M > 0.05 M_{\odot}$at distances $10-15$ AU from the
center is $50 \%$ higher than it would be without a companion by the
end of the simulation. Assuming a uniform gas-to-dust ratio in the
disk, the increase in surface density could compensate for the
vaporization of dust grains, making massive disks not less favourable
then light disks for giant planet formation via core accretion (see
above). Spiral arms might also gather solids as a result of pressure
gradients (Rice et al. 2004, 2006) leading to an enhanced gas-to-dust
ratio inside them, another effect that could favour core accretion.
Only more realistic calculations incorporating directly both
vaporization and the dynamical dust particles within the gaseous disk
will be able to settle this issue.

\subsection{Effects of artificial viscosity}

Boss (2006b) did not employ artificial viscosity in his standard
models, but did turn on the artificial viscosity in a subset of models
designed to determine to what extent the use of artificial viscosity
in either a finite-difference code (e.g., Pickett et al. 2000) or an
SPH code (e.g., Nelson 2000) might affect the disk instability
process. Boss (2006b) ran four models with varying amounts of
artificial viscosity, and found that only when the artificial
viscosity was set to a value a factor of 10 times higher than the
nominal value did the disk become so hot as to appreciably stifle
fragmentation.

One possible source of the different outcomes between the results of
Boss (2006b) and Nelson (2000) is the amount and effect of artificial
viscosity assumed in the two sets of models. Artificial viscosity
equivalent to an effective $\alpha$ viscosity with $\alpha = $ 0.002
to 0.005 was intentionally included in the Nelson (2000) models in
order to include the effects of shock heating. In the Boss (2006b)
models, artificial viscosity was not used, but the level of implicit
numerical viscosity appears to be equivalent to $\alpha \sim 10^{-4}$
(Boss 2004b), about 20 to 50 times lower than that in Nelson (2000).
Given the experiments of Boss (2006b) with artificial viscosity, the
use of this level of artificial viscosity in Nelson (2000) is
consistent with the absence of fragmentation and the difference in
cooling times in the two sets of models. Relatively short cooling
times are obtained in models without artificial viscosity ($\sim 1$ to
2 orbital periods, Boss 2004a), compared to the effective cooling time
obtained in Nelson (2000) of $\sim 5$ to $\sim 15$ orbital periods,
for orbital distances from 10 AU to 5 AU, respectively. 

The use of artificial viscosity by Nelson (2000) was motivated by a
good reason, namely to model shock dissipation in the disks, and
produced substantial heating in the disks. Nevertheless, an analysis
of the flux densities derived from his simulations fell nearly an
order of magnitude short of that required to reproduce the
observations of the L1551~IRS5 binary system, on which his initial
conditions were based. 

It is of some interest that Boss \& Yorke (1993, 1996) were able to
match spectral energy distributions of the T~Tauri system with
axisymmetric disk models similar to those that form the basis for Boss
(2006b) models, without using artificial viscosity. A part of the
contradiction can perhaps be explained by the fact that the system
modeled by Boss and Yorke was T~Tau, a system at a much later
evolutionary stage than L1551~IRS5, with correspondingly different
energy output. We look to future observations using the ALMA telescope
with great interest, because of the likelihood for observing younger,
and much more deeply embedded objects, of greater relevance to the
earliest stages of disk evolution where gravitational instabilities
are more probable. 

\subsection{Initial conditions in the context of star formation}

Are the initial conditions adopted in the existing simulations of
binary, self-gravitating protoplanetary disks realistic? In reality,
the two disks will be communicating since their beginnings, undergoing
mass transfer and growing out of gas flowing from the periphery of the
molecular cloud core. This is quite different from the setup assumed
in the simulations. Tidal perturbations and mass transfer might be too
sudden in the computations described in this chapter, while they will
be achieved gradually in reality. However, if star formation occurs in
gravoturbulent clouds, such as those modeled by Bate et al. (2002b),
rather than in isolated cores, disks will not have time to slowly
adjust to such an extremely dynamic environment by the time they
become gravitationally unstable. First of all, one way binary systems
can arise is via fragmentation of a bar-unstable core (Bate \& Burkert
1997). This occurs on the orbital  timescale of the rotating core , a
few hundred years to $10^3$ years, i.e. of the same order of the
binary orbital time considered in all the three papers discussed here,
rather than than on the collapse time of the cores, in the range
$10^3$ to $10^4$ years. The collapse of the individual cores would
occur on a timescale much smaller than the average collapse timescale
of the larger star-forming region. A short collapse time of cores is
suggested also by observations. Two examples are the observations that
prestellar cores have large enhancements in column densities and that
molecular abundances in them are consistent with a rapid collapse
(Aikawa 2004). The resulting systems would have undergone several
tidal interactions with bound or unbound companions since their birth.
Turbulent molecular clouds have velocity dispersions of order $2-5$
km$s^{1}$, which means the typical crossing time of a region $10^{-2}$
pc in size will be  $\sim 10^3$ years, The characteristic timescale of
encounters between cores in such a turbulent cloud has to be of the
same order of the latter timescale, i.e. once again comparable with
the binary orbital time. Nevertheless, Mayer et al. (2005) studied the
case of two self-gravitating disks reaching gradually the conditions
present at the begining of the simulations by starting with very
light, nearly non-self gravitating disks and growing the disk slowly
over the course of a few binary orbital periods (Figure 11). This way
the disk profile has time to adjust. The spiral arms tidally induced
on the third orbit are indeed slightly weaker than those in the
standard run, and transient localized overdensities are apparent that
were not present before, but no gravitationally bound clumps occur and
the outer disk temperature after one orbit ($\sim 300$ K) is
comparable to that in the original run 

\begin{figure}
\centering 
\includegraphics[height=7cm]{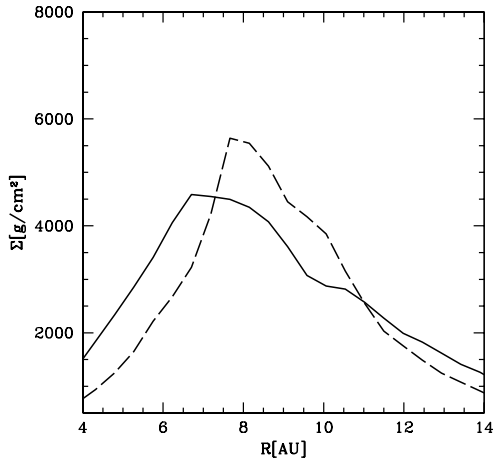}
\caption{Azimuthally averaged surface density profile after two binary
orbits for a run employing two massive disks with $M=0.1 M_{\odot}$.
The disks were evolved with a cooling time equivalent to $0.3$ the
local orbital time with self-gravity (solid line) and with
self-gravity  switched off on the first binary orbit (dashed line).}
\label{fig:1}       
\end{figure}

The final surface density profiles of the disks in the two runs are
also quite similar. Mass redistribution due to gravitational torques
leads to a profile which cannot be described by a single power-law,
has a remarkable density peak close to  $7-8$ AU, and is steeper than
$r^{-2}$ outside this radius (Figure 9,11). The surface density
profiles are steeper than those produced by gravitational
instabilities when there is no companion. The mass inflow towards the
center is greater. One would be tempted to conclude that the viscous
evolution of the disk, where the ``viscosity'' is due to gravitational
instability, is faster in binary systems. This might lead to a faster
dissipation of the disk and a faster growth of the star since gas
flows outside in. Some difficulty remains in establishing a solid
connection because it is not clear how the large, initially massless
hole in the middle of the disks, affects the mass transfer further
out.

In the simulations of Mayer et al. (2005) the restructuring of the
disk results from a combination of tidal torques and intrinsic
self-gravity (see section 4.5). Since in the early stages
protoplanetary disks should be massive enough to be self-gravitating
(Yorke \& Bodenheimer 1999), it seems that this profound restructuring
driven by the two simultaneous effects will likely occur in binary
systems, and will occur early.

New SPH simulations with a variable equation of state (Bate 2001) that
use a variable mass resolution technique to reach down to achieve a
spatial resolution of $\sim 0.1 AU$ in a rotating (non turbulent)
collapsing molecular core of a fraction of a parsec in siz do indeed
show substantial evolution of binary disks and mass inflow towards the
central star just as a result of self-gravity. The rapidly rotating
core collapses, becomes bar unstable and fragments into two clumps
that later become a pair of pre-stellar cores surrounded by a fairly
large, tidally truncated disk (about 30 or 80 AU in size), as shown in
Figure 12. The two systems have unequal masses; while each system
starts out with more than 2/3 of the mass being in the disk and the
rest in a dense central clump, the precursor of the star, less than
$0.1$ AU in size, after about two binary orbits (corresponding to
$\sim 2000$ years and to the time at which we stopped the simulation)
more than half of the disk mass has accreted onto the central clump.
At this stage disks are slightly lighter than the central clump.
Removal of angular momentum results from a combination of spiral
instabilities in the disks and tidal torques from the companion, with
te spiral arms being strengthened by the perturber as well. The
physical driver of accretion is just gravity in these simulations
since no other mechanisms to remove angular momentum are present
except gravitational torques (both intrinsic and tidal). The only
caveat is that artificial viscosity, while required to model physical
dissipation in shocks correctly, might also enhance angular momentum
transport. The accretion rate from the disk onto the central protostar
at the end of the simulation is nearly $5 \times 10^-5 M_{\odot}/yr$,
or about ten times higher than the accretion rate from the core onto
the disk, for the lightest, and thus the most tidally perturbed, among
the two systems, and about a factor of two lower for the other one.

\section{Conclusions}

We have noted that a number of numerical and physical effects can
either encourage or discourage a disk instability from forming
self-gravitating clumps that could become gas giant protoplanets.
There are indications that the artificial viscosity used in SPH codes
(Mayer et al. 2005, Nelson 2000) generally tends to suppress
fragmentation (Mayer et al. 2004) although the simulations of Nelson
(2000) had a lower artificial viscosity and yet did not find
fragmentation. However, when artificial viscosity is such that
simulations can match fluxes observed from protostellar systems, the
resulting high level of disk heating can prevent fragmentation. 
Using a gravitational softening length that is smaller than the SPH
smoothing length throughout most of the disk evolution, as in Mayer et
al. (2005), favors fragmentation, as does the discreteness noise in
SPH. Sharp disk edges promote fragmentation, while low resolution
(mass resolution in SPH, as set by the number of particles, or grid
size in grid codes) seems to suppress fragmentation (Mayer et al.
2004; Boss 2000), but not always, as seen by Nelson (2006), where it
enhanced fragmentation. The dependence on resolution is currently
being investigated systematically in an on-going code comparison that
involves both SPH and adaptive mesh refinement (AMR) codes;
preliminary results confirm an increasing susceptibility towards
fragmentation with increasing resolution (Mayer et al., in
preparation). A prevalence of low order modes (Nelson 2000; Mayer et
al. 2005; Boss 2006b) promotes fragmentation, as overdensities that
occur on larger scales are easier to resolve with finite resolution.
Finally, and perhaps most importantly, short cooling times promote
fragmentation (Mayer et al. 2005), while long cooling times prevent it
altogether (Nelson 2000).

Mayer et al. (2005) found that binary companions with semimajor axes
of 58 AU prevented disk fragmentation, unless the disks had moderate
masses ($0.05-0.08 M_{\odot}$) and cooled rapidly, indeed faster than
even simulations that appeal to convective cooling suggest (Boss 2002,
2003; Mayer et al. 2006). They also found that when the semimajor axis
was raised to 100 AU, fragmentation or transient clump formation
resulted in all cases studied. The results for large separations are
not in contrast with those of Boss (2006b). Instead, those for
separations of 58 AU are in disagreement with Boss (2006b), who found
fragmentation and transient clump formation even for semimajor axes of
both 50 AU, although even closer encounters (i.e., higher
eccentricities) tended to work against the formation of
self-gravitating clumps. The results of Mayer et al. (2000) are not in
contrast with those of Nelson (2000), although Nelson's disks were not
fragmenting even in isolation according to Nelson et al. (2000), which
does not allow to formulate the same conclusion reached by Mayer et
al. (2005) , namely that fragmentation is generally suppressed in
tight binary systems. Moreover, we note that most of the disks used by
Mayer et al. (2005) were marginally unstable by construction ($Q <
2$), and therefore the fragmentation seen for larger semi-major axes
might simply reflect the fact that as the separation increases the
results tend to converge to those for disks with no binary companion.
This in other words means that the fragmentation seen in such cases
could have nothing to do with the tidal perturbation of a companion.
Conversely, Boss uses mostly disks that would be stable in isolation
($Q \sim 2$ or larger), and hence the only logical outcome of his
calculations in presence of a companion is either that the disk remain
stable or that fragmentation is enhanced (i.e. no experiment was
constructed to see whether fragmentation could be suppressed). This
different logic behind the design of the simulations in the two works
complicates the comparison and calls for future attempts by these and
other authors to perform and compare exactly the same experiment.

\begin{figure}
\centering 
\includegraphics[height=10cm]{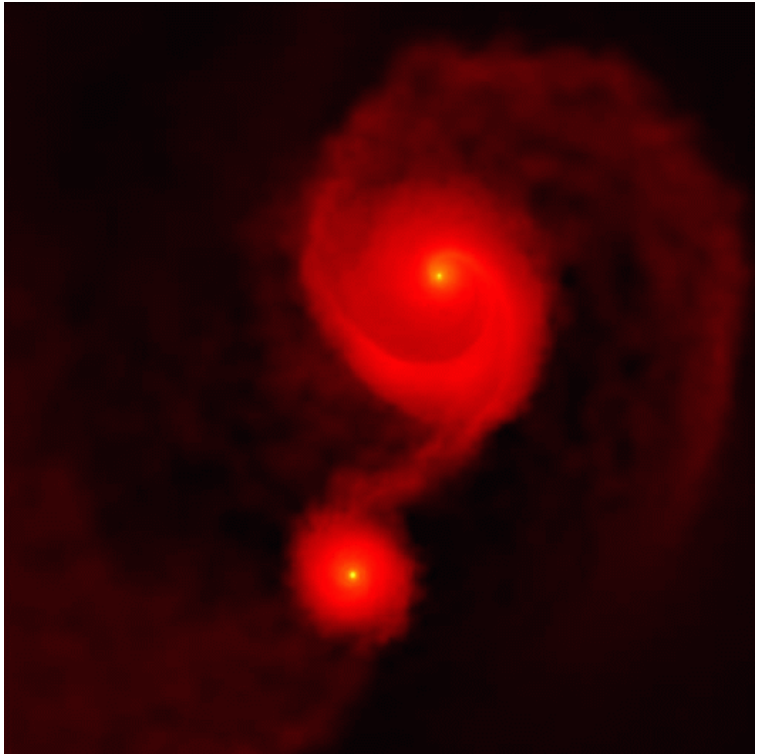}
\caption{Color coded density map of a binary protostar+disk system
resulting from the collapse of a rotating molecular cloud core with
initial density profile $\sim r^{-1}$. The box is 200 AU on a side,
and the system is shown about $5 \times 10^4$ years after the collapse
has been initiated, and a couple of binary orbits after the two disks
have formed from the fragmentation of a bar-unstable, rapidly rotating
protostellar core. The larger system weights $\sim 0.5 M_{\odot}$,
while the smaller system weights only $\sim 0.15 M_{\odot}$. The total
mass of the molecular core was $1 M_{\odot}$ and its size $10.000$ AU
at t=0. The simulation employs $500.000$ particles in total, but the
mass resolution in the inner 500 AU is higher than in the surrounding
volume, so that as many as $4/5$ of the particles are used only in
this inner region} \label{fig:1}       
\end{figure}

The results of Mayer et al. (2005) and Boss (2006b) suggest that the
formation of gas giant planets around binary stars with semimajor axes
of 100 AU or larger may be possible by the disk instability mechanism.
Note that for these systems core-accretion is in principle possible as
well since spiral shocks do not heat the gas to temperatures high
enough to vaporize major dust grain species (Mayer et al. 2005). For
smaller semimajor axes the situation is much more complex. For the
latter, Mayer et al. (2005) conclude that disk instability is unlikely
but that core-accretion might happen once the disk is light enough, $M
\sim 0.01 M_{\odot}$ (e.g. as a result of accretion onto the star)
that it can only form weak spiral shocks in which grains easily
survive. This is because in Mayer et al. (2005) the role of
self-gravity, and thus of disk mass, in the determining the strength
of the spiral shocks is crucial (see above). Nelson (2000) claimed
that both disk instability and core-accretion would be unlikely in
such systems, drawing the same conclusion that Mayer et al. (2005)
would have reached had they not considered binary systems composed of
light disks (but see above section 4.3 for possible caveats).
Finally, Boss (2006b) finds that these systems would fragment,
although he cannot follow the clumps for a long enough time to show
that they are long lasting. Despite the fact that disagreements exist
between the different works, it is clear from now that the tightest
binary systems might become an ideal testbed for theories of planet
formation. It is thus important for observers to refine their
estimates of the semimajor axes of binary systems containing gas giant
planets, in order to learn if these systems could have been formed by
disk instability. Post-formational orbital evolution of multiple
systems (e.g., decay of an unstable triple system) might be another
means to explain the observed binary systems with gas giant
companions. Finally, we recall that for intermediate semi-major axes
orbital eccentricities might also be important. Indeed, while Mayer et
al. (2005) adopt nearly circular orbits, Boss (2006b) chooses
eccentric orbits. Hence in Boss (2006b), for a given semi-major axis,
the disks will spend a larger fraction of the orbital time far away
from each other. The tidal perturbation will be more impulsive rather
than contiunous. In other fields of astrophysics which deal with
similar problems, such as in the study of galaxy interactions, it is
well known that impulsive or continuous tidal heating give rise to
quite different responses in a self-gravitating system, to the point
of determining a completely different structural evolution (Mayer et
al. 2001). What is seen in particular in the case of galaxies is that
impulsive encounters can generate ``cold'' features such as bars that
then survive for many orbital times, but the same features are erased
as the object increases too much its kinetic energy and/or thermal
energy owing to a continuous tidal perturbation such as that
associated with circular or nearly circular orbits. Similarly, one
could speculate that in encounters between disks on  eccentric orbits
transient overdensities might have a better chance to survive as the
same tidal force that triggered their formation fades away later
towards the apocenter of the the orbit. Again, simulations exploring a
larger parameter space are needed to assess if eccentricity is such an
important parameter and might help to partially reconcile the
disagreement between Boss (2006b) and Mayer et al. (2005). Disks
perturbed by fast-fly bies of other stars or brown dwarfs also suffer
significant tidal heating and do not fragment unless the cooling time
is very short (Lodato et al. 2007).

A major source of the differences obtained by Nelson (2000) and Mayer
et al. (2005) relative to Boss (2006b) could be the use of artificial
viscosity in the first two works based on SPH and its neglect by Boss
(2006b) The much longer cooling times in Nelson (2000) are also
expected to stifle fragmentation. Clearly, when artificial viscosity
is used to heat a disk, and this heat is unable to escape on an
orbital time scale, the chances for clump formation by disk
instability are severely reduced. This is a general issue for the disk
instability model, irrespective of the presence or not of a binary
companion. Some of the heating associated with artificial viscosity
will indeed a rise in nature as a result of turbulence and other
unresolved aspects of hydrodynamical flows. It remains for future work
to determine what is the proper amount of artificial viscosity to be
used in simulating realistic protoplanetary disks, and to determine
the proper boundary conditions at the surfaces of protoplanetary disk
that will allow the disk to radiate away energy at the correct rate.
The sensitivity to the cooling time is readily shown by the first in a
series of new simulations of binary disks with the algorithm for
radiative transfer described in Mayer et al. (2006). Intermediate mass
disks $0.05 M_{\odot}$ that were fragmenting in binaries in Mayer et
al. (2005), this being the only for which fragmentation was found even
for orbits with separations of $58$ AU at sufficiently short cooling
times, develop strong spiral arms  but no clumps when flux-limited
diffusion plus atmospheric cooling via blackbody radiation is used
(see Figure 13). This is not surprising since this latest radiation
physics model yields cooling times of order or slightly longer than
the orbital time (apparently via convection) while in Mayer et al.
(2005) these disks were fragmenting with shorter cooling times, half
or less than half the local orbital time. Indeed the average outer
disk temperatures (outside 5 AU) in these new simulation is larger
than $100$ K, in comparison with $60-70$ K in the corresponding run of
Mayer et al. (2005). The results of this simulation are in good
agreement with those of Nelson (2000).

\begin{figure}
\centering 
\includegraphics[height=6cm]{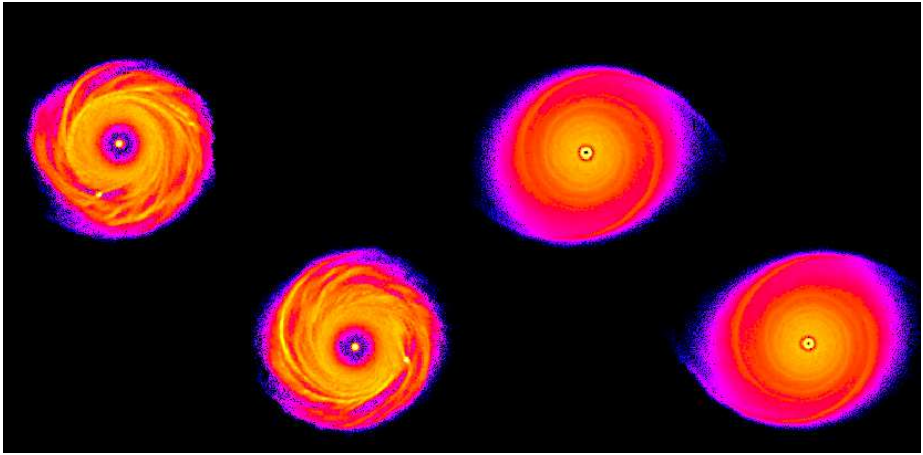}
\caption{Color coded density maps of two runs employing two disks with
masses $0.05 M_{\odot}$ moving on a binary orbit with average
separation of $60$ AU. The results are shown after $1.5$ binary
orbits. On the left a run in which the cooling time is fixed to $0.3$
orbital time is shown, on the right a newer run in which flux-limited
diffusion is employed and the disk cools at the surface as a
blackbody, resulting in a cooling time slightly longer than the
orbital time. Clump formation has occurred in the run with the short
cooling time while the in other run the two disks have achieved a
higher temperature, lower densities and a much weaker spiral
structure.} \label{fig:1}       
\end{figure}

However, the observational fact that binary stars with separations
small enough for mutual tidal interactions to be important are orbited
by gas giant planets means that somehow these planets can indeed form
even in these systems. Given the problems that core accretion
encounters as well in binary systems (Th\'ebault et al. 2004), disk
instability would seem to remain a possible means for forming gas
giants in binary systems.

\section{References}

Aikawa, Y., Proceedings of the Conference ``Star Formation at High Angular Resolution'', International Astronomical Union Symposium 221, XXV General Assembly of the IAU, Sydney, Australia, 22-25 July 2003. Eds: M. Burton, R. Jayawardhana, T. Bourke, 67, 2004

Alexander, ??? \& Ferguson, ??? 1994 ????

Bate, M. R. 2000, MNRAS, 314, 33 

Bate, M. R. \& Burkert, A. 1997, MNRAS, 228, 1060

Benz, W. 1990, in The Numerical Modeling of Stellar Pulsation, ed. J. R. 
Buchler (Dordrecht: Kluwer), 269 

Benz, W. 1991 ???

Benz, W. et al. 1990 ???

Boley, A. C. et al. 2006, ApJ, 651, 517

Boss, A. P. 1993, ApJ, 417, 351
 
Boss, A. P. 1996, ApJ, 469, 906

Boss, A. P. 1997, Science, 276, 1836

Boss, A. P. 1998, ApJ, 503, 923

Boss, A. P. 2000, ApJ, 545, L61

Boss, A. P. 2001, ApJ, 563, 367

Boss, A. P. 2002, ApJ, 576, 462
 
Boss, A. P. 2003, ApJ, 599, 577

Boss, A. P. 2004a, ApJ, 610, 456

Boss, A. P. 2004b, ApJ, 616, 1265

Boss, A. P. 2005, ApJ, 629, 535

Boss, A. P. 2006a, ApJ, 637, L137

Boss, A. P. 2006b, ApJ, 641, 1148

Boss, A. P. \& Myhill, E. A. 1992, ApJS, 83, 311

Boss, A. P. \& Yorke, H. W. 1993, ApJ, 411, L99

Boss, A. P. \& Yorke, H. W. 1996, ApJ, 469, 366

Cai, K. et al. 2006a, ApJ, 636, L149

Cai, K. et al. 2006b, ApJ, 624, L173 (Erratum)

Cameron, A. G. W. 1978, Moon Planets, 18, 5

Chauvin, G. et al. 2006, A\&A, 456, 1165

D'Alessio, P., Calvet, N., \& Hartmann, L., 2001, ApJ, 553, 321

Duquennoy, A. \& Mayor, M. 1991, A\&A, 248, 485

Durisen, R. H., Cai, K., Mej\'ia, A. C. \& Pickett, M. K. 2005, Icarus,
173, 417

Durisen, R. H., et al. 2007, in Protostars \& Planets V, ed. B. Reipurth
(Tucson: Univ. Arizona Press), in press

Eggenberger, A., Udry, S. \& Mayor, M. 2004, A\&A, 417, 353 

Evrard, A.E. 1990, ApJ, 363, 349

Gammie, C. F. 2001, ApJ, 553, 174

Gingold, R. A. \& Monaghan, J. J. 1977, MNRAS, 181, 375 

Goldreich. P. \& Lynden-Bell, D. 1965, ???

Herant, M. \& Woosley, S. 1994, ApJ, 425, 814

Johnson, B. \& Gammie, C. F. 2003, 597, 131

Kawakita, H., et al. 2001, Science, 294, 1089

Kuiper, G. P. 1951, Proc. Nat. Acad. Sci., 37, 1

Laughlin, G., Korchagin, V. \& Adams, F. C. 1997, ApJ, 477, 410

Lin, D. N. C. \& Pringle, J. E. 1987, MNRAS, 225, 607 

Lodato, G. \& Rice, W. K. M. 2004, MNRAS, 351, 630 

Lodato, G., Meru, F. Clarke, C. \& Rice, W. K. M. 2007, MNRAS, in press

Lucy, L. B. 1976 ???

Mayer, L., Colpi, M., Governato, F., Moore, B., Quinn, T., Stadel, J., \& Lake, G., 2001, ApJ, 559, 754

Mayer, L., Quinn, T., Wadsley, J. \& Stadel, J. 2002, Science, 298, 1756

Mayer, L., Quinn, T., Wadsley, J. \& Stadel, J. 2004, ApJ, 609, 1045

Mayer, L., Wadsley, J., Quinn, T. \& Stadel, J. 2005, MNRAS, 363, 641

Mayer, L., Lufkin, G., Quinn, T. \& Wadsley, J., 2007, ApJ, submitted

Mej\'ia, A. C., Durisen, R. H., Pickett, M. K. \& Cai, K., 2005, ApJ, 
619, 1098

Monaghan, J. J. 1992, ARAA, 30, 543 

Morris, J. P., Monaghan, J. J., 1997, J. Comp. Phys., 136, 41

Nelson, A. F. 2000, ApJ, 537, L65

Nelson, A. F. 2006, MNRAS, 373, 1039-1073

Nelson, A. F., Benz, W., Adams, F. C. \& Arnett, W. D. 1998, ApJ, 502, 342

Nelson, A. F., Benz, W. \& Ruzmaikina, T. V. 2000, ApJ, 529, 357

Paczy\'nski, B. 1978, ???

Papaloizou, J. C. \& Savonije, G. J. 1991, MNRAS, 248, 353

Patience, J., et al. 2002, ApJ, 581, 654 

Pickett, B. K., Durisen, R. H. \& Davis, G. A. 1996, ApJ, 458, 714

Pickett, B. K., Durisen, R. H. \& Link, R. 1997, Icarus, 126, 243
 
Pickett, B. K., Cassen, P., Durisen, R. H. \& Link, R. 1998, ApJ, 504, 468

Pickett, B. K., et al. 2000, ApJ, 529, 1034

Pickett, B. K., et al. 2003, ApJ, 590, 1060

Pollack, J. B., McKay, C. \& Christofferson,  B. M., 1985, Icarus, 64, 471

Pollack, J.B., Hollenbach, D., Beckwith, S., Simonelli, D.P., Roush, T. \& Fong, W. 1994, ApJ, 421, 615

Rafikov, R. R. 2006, ApJ, submitted (astro-ph/0609549)

Raghavan, D. et al. 2006, ApJ, 646, 523

Rice, W. K. M., Armitage, P. J., Bate, M. R. \& Bonnell, I. A. 2003a,
MNRAS, 339, 1025

Rice, W. K. M. et al. 2003b, MNRAS, 346, L36

Springel, V. \& Hernquist, L. 2002, MNRAS, 333, 649

Th\'ebault, P., et al. 2004, A\&A, 427, 1097
 
Tomley, L., Cassen, P. \& Steiman-Cameron, T. 1991, ApJ, 382, 530

Tomley, L., Steiman-Cameron, T. Y. \& Cassen, P. 1994, ApJ, 422, 850

Toomre, A. 1964, ApJ, 139, 1217

Udry, S. Eggenberger, A., Beuzit, J.L., Lagrange, A.M., Mayor,
  M., \& Chauvin, G., 2004, in ``The Environment and Evolution of Double and Multiple Stars'', Proceedings of IAU Colloquium 191, held 3-7 February, 2002 in Merida, Yucatan, Mexico. Edited by Christine Allen \& Colin Scarfe. Revista Mexicana de Astronomi'a y Astrofi'sica (Serie de Conferencias) Vol. 21, pp. 215-216 (2004)

Wadsley, J., Stadel, J., \& Quinn, T.R.,  New Astronomy, 9, 137

Yorke, H. W. \& Bodenheimer, P. 1999, ???

\end{document}